\documentclass[article]{aa}

\bibpunct{(}{)}{;}{a}{}{,}

\usepackage{graphicx}
\usepackage{tabularx}
\usepackage{txfonts}
\usepackage{textcomp}
\usepackage{gensymb}
\usepackage{url}
\usepackage{footmisc}
\usepackage{amsmath}
\usepackage{fixltx2e}
\usepackage{hyperref}
\usepackage{animate}
\usepackage[author={Karsten Schindler}]{pdfcomment}
\usepackage{xcolor}

\begin{document} 

   \title{Results from a triple chord stellar occultation and\\far-infrared photometry\thanks{\textit{Herschel} is an ESA space observatory with science instruments provided by European-led Principal Investigator consortia and with important participation from NASA.} of the trans-Neptunian object (229762)~2007~UK\textsubscript{126}}

   \author{K. Schindler
          \inst{1,2}
          \and
          J. Wolf
          \inst{1,2}
          \and 
          J. Bardecker
          \inst{3}\fnmsep\thanks{Also affiliated with the Western Nevada Astronomical Society, 2699 Van Patten Avenue, Carson City, NV 89703, USA and the Research and Education Collaborative Occultation Network (RECON).}
          \and
          A. Olsen
          \inst{3}
        \and
          T. M\"uller
          \inst{4}
        \and
          C. Kiss
          \inst{5}
        \and
          J. L. Ortiz
          \inst{6}
        \and
        \\ F. Braga-Ribas
        \inst{7,8}        
        \and
        J. I. B. Camargo
        \inst{7,9}
        \and
        D. Herald
        \inst{3}
        \and
        A. Krabbe
        \inst{1}
        }

   \institute{Deutsches SOFIA Institut, Universit\"at Stuttgart,
              Pfaffenwaldring 29, 70569 Stuttgart, Germany\\
              \email{schindler@dsi.uni-stuttgart.de}
         \and
             SOFIA Science Center, NASA Ames Research Center, 
          Mail Stop N211-1, Moffett Field, CA 94035, USA
          \and
             International Occultation Timing Association (IOTA)
        \and
          Max Planck Institute for Extraterrestrial Physics, 
          Giessenbachstrasse 1, 85748 Garching, Germany
        \and
        Konkoly Observatory, Research Centre for Astronomy and Earth Sciences, Hungarian Academy of Sciences,
        Konkoly Thege 15-17, 1121 Budapest, Hungary
        \and
        Instituto de Astrofisica de Andalucia-CSIC, Glorieta de la Astronomia, 3,
        18080 Granada, Spain
        \and
        Observat\'orio Nacional/MCTI, Rua Gal. Jos\'e Cristino 77, Rio de Janeiro, RJ 20921-400, Brazil 
        \and
	Federal University of Technology - Paran\'a (UTFPR / DAFIS), Rua Sete de Setembro, 3165, CEP 80230-901, Curitiba, PR, Brazil
	\and
        Laborat\'orio Interinstitucional de e-Astronomia - LIneA, Rua Gal. Jos\'e Cristino 77, Rio de Janeiro, RJ 20921-400, Brazil
           }

   \date{Received 01 April 2016; revised 10 August 2016; accepted 10 October 2016}


  \abstract
   {A stellar occultation by a trans-Neptunian object (TNO) provides an opportunity to probe the size and shape of these distant solar system bodies. In the past seven years, several occultations by TNOs have been observed, but mostly from a single location. Only very few TNOs have been sampled simultaneously from multiple locations. Sufficient data that enable a robust estimation of shadow size through an ellipse fit could only be obtained for two objects.}
   {We present the first observation of an occultation by the TNO 2007~UK\textsubscript{126} on 15 November 2014, measured by three observers, one nearly on and two almost symmetrical to the shadow's centerline. This is the first multi-chord dataset obtained for a so-called detached object, a TNO subgroup with perihelion distances so large that the giant planets have likely not perturbed their orbits. We also revisit \textit{Herschel}/PACS far-infrared data, applying a new reduction method to improve the accuracy of the measured fluxes. Combining both datasets allows us to comprehensively characterize 2007~UK\textsubscript{126}.}
  {We use error-in-variable regression to solve the non-linear problem of propagating timing errors into uncertainties of the ellipse parameters. Based on the shadow's size and a previously reported rotation period, we expect a shape of a Maclaurin spheroid and derive a geometrically plausible size range. To refine our size estimate of 2007~UK\textsubscript{126}, we model its thermal emission using a thermophysical model code. We conduct a parametric study to predict far-infrared fluxes and compare them to the \textit{Herschel}/PACS measurements.} 
   {The favorable geometry of our occultation chords, combined with minimal dead-time imaging, and precise GPS time measurements, allow for an accurate estimation of the shadow size (best-fitting ellipse with axes $645.80 \pm 5.68~\mathrm{km} \times 597.81 \pm 12.74~\mathrm{km}$) and the visual geometric albedo ($p_\mathrm{V} = 15.0 \pm 1.6 \%$). By combining our analyses of the occultation and the far-infrared data, we can constrain the effective diameter of 2007~UK\textsubscript{126} to $d_\mathrm{eff} = 599 - 629~\mathrm{km}$. We conclude that subsolar surface temperatures are in the order of $\approx~50 - 55~\mathrm{K}$.}
{}

   \keywords{Occultations -- 
                Kuiper belt objects: individual: (229762) 2007 UK126 --
                        Radiation mechanisms: thermal -- 
                        Methods: data analysis
               }

  \titlerunning{Results from an occultation and FIR photometry of (229762)~2007~UK\textsubscript{126}}
  \authorrunning{Schindler et al.}
   \maketitle


\section{Introduction}

Stellar occultations are the best opportunity to directly and accurately determine the size and shape of a solar system body remotely from Earth. They allow improvements of orbital elements and have the potential to discover previously unknown satellites \citep[e.g.][]{Timerson2013,Descamps2011}, or even ring systems, as recently revealed for Centaur (10199)~Chariklo \citep{Braga-Ribas2014a}. The shape of the light curve can reveal the presence of an atmosphere, and enable the study of its properties \citep[e.g.][]{Person2013}. While observations of occultations only require basic photometric tools, the technical challenge is to acquire images at high cadence with little-to-no dead time between acquisitions, and with very precise timing information to measure disappearance and reappearance times of the occulted star with little uncertainty. Knowledge of the target's apparent orbital velocity and length of the occultation allow for a direct calculation of its size at the location of the sampled plane or chord. Multiple observers distributed across the shadow path can sample the occulting object at different locations, which provides information on its shape. 

Trans-Neptunian objects (TNOs) are considered the most pristine objects in our solar system. Being left-overs from the very early stage of the accretion phase, estimates of their sizes, densities and albedos, as well as constraints on their composition can provide critical information on the evolution of the solar system. Given their large geocentric distance and very small angular diameter, the prediction of a TNO's shadow path on the Earth's surface during an occultation is difficult. Owing to the very long orbital periods of TNOs, only a very small fraction of their orbits has been observed since their discovery, leaving relatively large uncertainties in their orbital elements. Another issue for shadow path predictions are astrometric uncertainties of currently available star catalogs, which are inherited by orbit and ephemeris calculations for TNOs and which blur the true position of the occulted star. Besides random astrometric uncertainties that typically increase towards fainter stars, catalogs usually have zonal errors. Another dominating error that can shift a shadow path significantly is potential stellar duplicity, which usually cannot be excluded in advance. We also have no information on albedo variations on the TNO's surface, which could cause a periodic shift of the center of light. All these error sources make predictions and successful observation campaigns for TNO occultations a difficult task \citep{Bosh2016}. To improve shadow path predictions, extensive astrometric observations with high precision of both the target star and the occulting body are usually conducted for many weeks in advance of an event. The first data release of the GAIA star catalog \citep{Lindegren2016,GAIA2016} and all subsequent data releases\footnote{\url{http://www.cosmos.esa.int/web/gaia/release}} up to the final GAIA catalog in 2022 are expected to improve the accuracy of occultation predictions significantly, since GAIA will offer the most accurate astrometry ever obtained. GAIA will also significantly contribute towards stellar duplicity measurements.

A number of occultation events covering at least 11~different TNOs \citep[see e.g.][]{Ortiz2014,Braga-Ribas2014b} have been successfully observed to date, not counting the well-studied Pluto system. However, most occultations by TNOs could only be observed at a single location, while the acquisition of multiple chords during an event has been extremely rare. Table~\ref{tab:table1} gives an overview of all successful multi-chord observations of occultations by TNOs (except Pluto) that have been published in peer-reviewed literature so far, covering only four objects to date. (50000)~Quaoar \citep{Braga-Ribas2013} is currently the best sampled TNO, having five unique chords (i.e. chords that are geographically sufficiently distant to each other to sample the occulting object at distinguishable locations) that were recorded during a single event (04~May~2011). In this dataset, the time resolution and accuracy of the two chords above the centerline have been low, resulting in large uncertainties of ingress and egress times and consequently in some ambiguity of the ellipse fit. (136472)~Makemake \citep{Ortiz2012} has been generally sampled with good time resolution, but three of the four unique chords were very close to each other and almost located at the centerline of the ellipse fit, while the fourth chord was located below the centerline, and no chord was located above, likewise resulting in ambiguity of the shadow geometry. For the remaining two objects, (136199)~Eris and (55636)~2002~TX\textsubscript{300}, only two unique chords are available, which results in an under-determined ellipse fit. To our knowledge, multi-chord events were also recorded for the following objects, but have not been published in peer-reviewed literature so far: (208996)~2003~AZ\textsubscript{84}, (20000)~Varuna and (84922)~2003~VS\textsubscript{2}.

\begin{table*}[tbp]
  \setlength{\extrarowheight}{2.5pt}
  \caption{Multi-chord observations of stellar occultations by trans-Neptunian objects (TNOs) published in peer-reviewed literature to date.}
  \centering
  \label{tab:table1}
  \begin{tabular}{c >{\centering}p{1.7cm} >{\centering}p{3.2cm} >{\centering}p{3.2cm} >{\centering}p{3.2cm} >{\centering\arraybackslash}p{1.6cm}}
    \hline\hline
    Object & \mbox{Dynamical}\linebreak\mbox{class} & Date (UTC) & Location & \mbox{No. of recorded chords}\linebreak\mbox{(No. of unique chords)} & Key\linebreak reference \\
    \hline
    (55636) 2002 TX\textsubscript{300} & HC & 09 October 2009 & Hawaii & 2 & 1 \\
    (136199) Eris & SDO & 06 November 2010 & Chile & 3 (2) & 2 \\
    (136472) Makemake & HC & 23 April 2011 & Chile, Brazil & 7 (4) & 3 \\
    (50000) Quaoar & HC & \mbox{04 May 2011}\linebreak\mbox{17 February 2012} & \mbox{Chile, Brazil, Uruguay}\linebreak\mbox {France, Switzerland} & \mbox{6 (5)}\linebreak\mbox{4 (2)} & 4 \\
    (229762) 2007 UK\textsubscript{126} & DO & \mbox{15 November 2014} & \mbox{USA} & 3 & this work \\
    \hline
    \end{tabular}
    \tablefoot{Numbers in brackets indicate the number of chords that were geographically sufficiently distant to each other to sample the occulting object at distinguishable (i.e. unique) locations. Abbreviations: SDO - Scattered disk object; HC - Hot classical Kuiper belt object; DO - Detached Object.}
    \tablebib{
        (1) \citet{Elliot2010}; (2) \citet{Sicardy2011}; (3) \citet{Ortiz2012}; (4) \citet{Braga-Ribas2013}}
\end{table*}

In this paper, we report results from the very first multi-chord observation of a stellar occultation by a TNO that is classified as a detached object according to the widely used definitions by \citet{Gladman2008}. This term denotes a sub class of TNOs with perihelion distances so large that Neptune and the other giant planets have likely not perturbed their orbits in the past. In addition to being the first comprehensive study of an object of this dynamical class, the dataset presented in this paper is also extremely rare compared to previously achieved observations of occultations by TNOs: To our knowledge, {it} is only the second after (50000) Quaoar that provides a sufficient number of chords (at least three chords are required for an ellipse fit) that are well spaced (in the present case sampling the target simultaneously above, very close, and below the centerline by extremely favorably distributed, quasi symmetrically located observers). The size derived from the occultation measurements is used as an important new constraint for thermophysical modeling based on \textit{Herschel} far-infrared (FIR) flux measurements. This combined analysis allows us to comprehensively characterize 2007~UK\textsubscript{126}. 


\section{Known properties of 2007~UK\textsubscript{126} from literature} \label{sec:properties}

2007~UK\textsubscript{126} was discovered at Palomar Observatory on 19~October~2007. According to the database of the Minor Planet Center~(MPC)\footnote{\url{http://www.minorplanetcenter.net}}, it has been identified subsequently on older images taken at Siding Spring Observatory and Palomar dating back to August~1982, which allowed improvements of orbit calculations ($i = 23.34\degree$, $e = 0.49$, $a = 74.01~\mathrm{AU}$). Having an orbital period of $636.73~\mathrm{yr}$, the object is approaching perihelion (at a heliocentric distance of $r_\mathrm{H} = 37.522~\mathrm{AU}$), which it will pass on 18~March~2046, based on current ephemeris data provided by the JPL Small-Body Database\footnote{\url{http://ssd.jpl.nasa.gov/sbdb.cgi} | \url{http://ssd.jpl.nasa.gov/horizons.cgi}}. Around perihelion, the target will have its largest apparent magnitude of $m_\mathrm{V} \approx 19.13~\mathrm{mag}$ (in V-band) during its entire orbit according to MPC estimates.

Photometric and spectroscopic data of 2007~UK\textsubscript{126} were acquired with several instruments at ESO's~VLT in a coordinated campaign on 21~and~22~September~2008. \citet{Perna2010} report visible and infrared photometry (VRIJH) taken with VLT/FORS2 and VLT/ISAAC. They estimated the absolute magnitude (defined as the apparent magnitude at a distance of 1~AU both to the Sun and to the observer at zero phase angle) of 2007~UK\textsubscript{126} in \mbox{V-band} as $H_\mathrm{V} = 3.69 \pm 0.04~\mathrm{mag}$. Their \mbox{R-band} measurement allows us to derive $H_\mathrm{R} = 3.07 \pm 0.04~\mathrm{mag}$. While these photometric measurement uncertainties $\mathrm{\Delta} H_\mathrm{phot}$ result from noise in the data, they do not consider possible periodic magnitude variations $\mathrm{\Delta} m$ owing to the rotation of the body. To take these into account as well, \citet{Santos-Sanz2012} proposed an additional error term of $\mathrm{\Delta} H_\mathrm{rot} = 0.88 \cdot \frac{\mathrm{\Delta} m}{2}$. This is justified by assuming a sinusoidal light curve, where roughly 88\% of its amplitude contains 68.3\% of the function values. As the light curve amplitude was not measured back in 2012, a peak-to-peak amplitude of $\mathrm{\Delta} m = 0.2~\mathrm{mag}$ was assumed, arguing that roughly 70\% of a sample of TNO light curves studied by \citet{Duffard2009} showed less magnitude variations. Considering both independent error sources, the total measurement uncertainty becomes $\mathrm{\Delta} H = \sqrt{\left( \Delta H_\mathrm{phot} \right)^2 + \left( \Delta H_\mathrm{rot} \right)^2}$. This leads to refined absolute magnitude estimates of $H_\mathrm{V} = 3.69 \pm 0.10~\mathrm{mag}$ and $H_\mathrm{R} = 3.07 \pm 0.10~\mathrm{mag}$.

\begin{figure}[tb]
   \centering
   \includegraphics{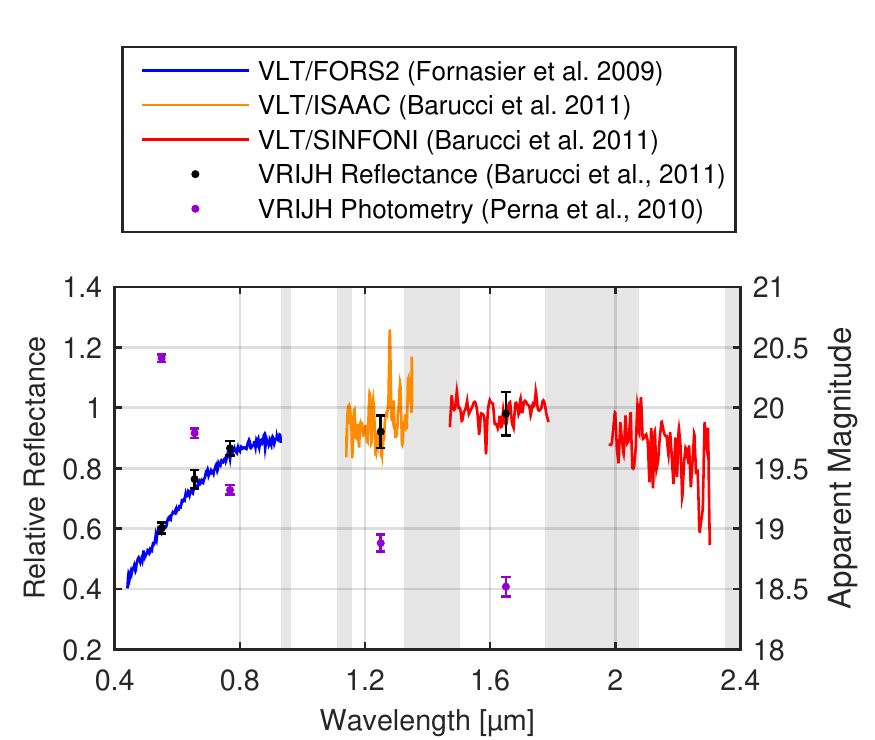}
      \caption{Visible and near-infrared reflectance spectra and photometry, obtained using multiple instruments at ESO's VLT on 21 and 22 September 2008. To concatenate spectral data from different instruments, photometry (purple) was converted to relative reflectance (black) and used as an indicator for alignment \citep[see discussion in][]{Barucci2011}. Zones marked in gray denote telluric bands of the Earth's atmosphere. See text for details.}
   \label{fig:figure1}
\end{figure}

Interestingly, \citet{Perna2010} note that 2007~UK\textsubscript{126} was the only object in their study that they were unable to categorize into any of the four TNO taxonomic classes. By comparing color indices (B-V, V-R, V-I, V-J, V-H, V-K) to those of the Sun, a TNO can be classified as BB (neutral, or ``blue'' in color), BR (intermediate ``blue-red''), IR (moderately ``red'') and RR (very ``red''); details are given in \citet{Barucci2005} and \citet{Fulchignoni2008}. While visible photometry provides classifications of 2007~UK\textsubscript{126} both as IR or RR, infrared photometry allows classifications both as BB or BR. This could indicate that the current taxonomic scheme needs to be refined in the future. \citet{Fornasier2009} report visible spectroscopy of 2007~UK\textsubscript{126} obtained with VLT/FORS2 (0.44~--~0.93~\textmu m). The spectral continuum between 500~--~800~nm has a slope of $(19.6 \pm 0.7)\%$~per~100~nm, which is considered moderately red and well within the slope range of various samples of the TNO population \citep[see e.g.][]{Lacerda2014}. \citet{Barucci2011} report spectroscopy performed with VLT/ISAAC \mbox{(1.1~--~1.4~\textmu m)} and VLT/SINFONI \mbox{(1.49~--~2.4~\textmu m)}. They calculated the relative flux difference D between \mbox{1.71~--~1.79~\textmu m} and \mbox{2.0~--~2.1~\textmu m,} which could hint at a possible water ice feature around 2.0~\textmu m. Owing to the limited signal-to-noise ratio (S/N) of the data, a presence of this type of feature could not be confirmed with sufficient statistical significance ($D = (11 \pm 7) \%$). Using Hapke's radiative transfer model, a synthetic spectrum was calculated by considering various compounds. The best Hapke model fit is based on a geometric albedo (in V-band) of $p_\mathrm{V} = 0.20$ (a value that we disprove in our subsequent analysis) and results in an estimated surface composition of 12\%~amorphous water ice, 20\%~kaolinite, 17\%~Titan tholins, 32\%~Triton tholins, 4\%~kerogen and 15\%~carbon. However, the authors conclude that ``much of the information obtained from spectral modeling is nonunique, especially if the albedo is not available, the SNR is not very high and / or there are no specific features of particular components'', so these findings need to be viewed with caution. For a full discussion of this spectral analysis, see \citet{Barucci2011}. All spectral data available from literature is summarized in Figure~\ref{fig:figure1}. We conclude that the spectral data available at this point cannot reliably constrain surface composition. 

Estimating a size from the magnitude of an object is impossible without the knowledge of its geometric albedo, as a small object with a highly reflective surface cannot be distinguished from a large object with a dark surface. The thermal emission of the target can be used as a second constraint to overcome this ambiguity. \citet{Santos-Sanz2012} report results derived from far-infrared photometry in three bands (\mbox{60~--~85~\textmu m}, \mbox{85~--~120~\textmu m}, \mbox{130~--~210~\textmu m}) using \textit{Herschel}/PACS. By applying a hybrid standard thermal model, they conclude that 2007~UK\textsubscript{126} has a diameter in the range of $d = 599 \pm 77~\mathrm{km}$, a geometric albedo (\mbox{in V-band}) of $p_\mathrm{V} = 16.7^{+5.8}_{-3.8}\%$, and a beaming factor of $\eta = 1.20 \pm 0.35$, a factor that is empirically found to scale modeled surface temperature to observations \citep[see discussion in][]{Mueller2009}. \citet{Santos-Sanz2012} also conclude that a beaming factor close to $\eta = 1$ indicates ``very low thermal inertia and/or very large surface roughness'', which could imply a highly porous surface with very low thermal conductivity as a possible scenario. The derived size implies that the body is a dwarf planet candidate that could have formed into a regular, round shape due to its own gravity. This depends on the density and material strength of its compounds and its rotation period. For incompressible fluids, figures of hydrostatic equilibrium have been derived, of which two are of particular importance: A Maclaurin spheroid (an oblate spheroid with axes $a=b>c$) and a Jacobi ellipsoid \citep[all three axes $a>b>c$ having different length,][]{Chandrasekhar1967}. However, the discussion of these shapes only represents a theoretical limiting case, since bodies made of solid matter have mechanical strength. As pointed out by \citet{Sheppard2002}, a fractured interior as a result of impacts could lead to fluid-like behavior, but conditions are never determined solely by hydrodynamics.

\citet{Thirouin2014} report photometric light curves taken at the Telescopio Nazionale Galileo (TNG) over a time span of $\approx~10~\mathrm{h}$ in total, distributed over three nights in October~2011. The obtained light curve of 2007~UK\textsubscript{126} is very flat and varies by just $\mathrm{\Delta} m = 0.03 \pm 0.01~\mathrm{mag}$ peak-to-peak. The light curve amplitude depends on the shape of the body, the viewing geometry and albedo variation across the surface, all three being unknown factors. Both a single-peaked (Maclaurin spheroid with albedo inhomogeneity) and double-peaked (Jacobi ellipsoid) light curve are plausible. \citet{Thirouin2014} have used a criterion of $\mathrm{\Delta} m = 0.15~\mathrm{mag}$ peak-to-peak amplitude to distinguish between albedo and shape-related light curve variations; consequently, this object was analyzed assuming a single-peak solution. Still, they were unable to estimate a secure sidereal rotation period $P$, since data indicate multiple possible solutions of $P = \{11.05~\mathrm{h}, 14.30~\mathrm{h}, 20.25~\mathrm{h}\}$, giving the $P = 11.05~\mathrm{h}$ solution only a minimal higher likelihood compared to the other two. They conclude that their data only allow the rotation period to be constrained with a lower limit of $P > 8~\mathrm{h}$ at this time. As a rough first indicator, \citet{Thirouin2014} also derived a lower density limit of $\rho > 0.32~\mathrm{g~cm^{-3}}$ based on the assumption of a homogeneous Jacobi ellipsoid in hydrostatic equilibrium that rotates in $P = 8~\mathrm{h}$. They note that this assumption on shape is in conflict with the preference for an oblate spheroid, and emphasize that this estimate is very vague and can be unrealistic. \citet{Thirouin2014} also state an absolute magnitude of $H = 3.4~\mathrm{mag}$ without specifying a filter band (presumably R) or uncertainty estimate, and indicate that no absolute photometry has been derived; hence we discarded their value. As we now have additional information on the peak-to-peak amplitude of the light curve, we can reapply the approach from \citet{Santos-Sanz2012}, as explained earlier, to derive an improved uncertainty estimate of their absolute magnitude measurements. We obtain $\mathrm{\Delta} H = \sqrt{\left(0.04~\mathrm{mag}\right)^2 + \left(0.88 \cdot \frac{0.04~\mathrm{mag}}{2} \right)^2} = 0.044~\mathrm{mag}$. 

\textit{Hubble Space Telescope (HST)}/\mbox{WFPC2-PC} observations on 13~November~2008\footnote{\url{http://www2.lowell.edu/users/grundy/tnbs/229762_2007_UK126.html}} have revealed the existence of a satellite \citep{Noll2009,Santos-Sanz2012}, but its orbit could not be determined so far. The satellite has a magnitude difference of $\Delta m = 3.79 \pm 0.24~\mathrm{mag}$, measured in the \mbox{\textit{HST}/WFPC2-PC} F606W band ($599.7~\mathrm{nm} \pm 75~\mathrm{nm}$, see Mikulski Archive for Space Telescopes for more details\footnote{\url{https://archive.stsci.edu/}}). %


\section{Observations}

\subsection{Visual observations of the stellar occultation}\label{sec:observations}
2007~UK\textsubscript{126} occulted USNO~CCD~Astrograph~Catalog~4 star \mbox{UCAC4~448-006503} in the constellation Eridanus ($\mathrm{RA_{J2000}} = 04\mathrm{h}~29\mathrm{m}~30.6\mathrm{s}$, $\mathrm{Dec_{J2000}} = -00\degree~28\arcmin~20.9\arcsec$, apparent magnitude $m_\mathrm{V}~=~15.86~\mathrm{mag}$ \mbox{(V-band)}, $m_\mathrm{B}~=~17.00~\mathrm{mag}$ \mbox{(B-band)}, $m_\mathrm{J}~=~14.34~\mathrm{mag}$ \mbox{(J-band)}\footnote{\url{http://vizier.u-strasbg.fr/viz-bin/VizieR-5?-ref=VIZ5599d6063ad7&-out.add=.&-source=I/322A/out&UCAC4===448-006503}}) on 15~November~2014~UTC. At the time of the occultation, the target had a geocentric distance of $r_\mathrm{G} = 42.572~\mathrm{AU}$ and a heliocentric distance of $r_\mathrm{H} = 43.47~\mathrm{AU}$. It was close to opposition (01~December~2014, phase angle $\chi = 0.56$) and had an approximate apparent visual magnitude of $m_\mathrm{V}~\approx~19.84~\mathrm{mag}$. Details on the prediction of this event can be found in \citet{Camargo2014}.

The event was announced in October~2014 via the RIO TNO Events data feed\footnote{Maintained by D.~Gault, Australia, publishing predictions from the RIO TNO Group (with members from Observat\'orio Nacional/MCTI and Observat\'orio do Valongo/UFRJ, Rio de Janeiro, Brazil; and Observatoire de Paris-Meudon/LESIA, Meudon, France).} in OccultWatcher\footnote{\url{www.occultwatcher.com}}, a software program written by H.~Pavlov \citep{Pavlov2014} that allows coordination among professional and amateur observers, planning of campaigns, and deployment of mobile setups across the predicted path to acquire as many chords as possible. Before the event, 25~stations registered to attempt an observation of the event, while 16~stations reported back afterwards. In addition, this event was announced in a private email by J.~L.~Ortiz.

Unfortunately, weather conditions clouded out most observers in California and the south western U.S.. Three observers reported a successful observation of the event on OccultWatcher. Table~\ref{tab:table2} provides an overview of their locations and setups. It turned out that all successful observers were coincidentally positioned with almost equal spacing to each other in relation to the shadow path, sampling the shadow almost symmetrically to the centerline. Schindler~\&~Wolf observed at about 89~--~91\% humidity and low transparency sky conditions in California; fortunately, the event happened during a larger gap in a high cirrus cloud cover seen around the time of the event. The 60~cm Astronomical Telescope of the University of Stuttgart (ATUS)\footnote{\url{https://www.dsi.uni-stuttgart.de/forschung/atus.html}}, located at Sierra Remote Observatories about an hour's drive north-east of Fresno, was controlled remotely from Germany via internet connection. Bardecker, located in Nevada, also reported that high cirrus clouds moved in about two minutes after the event. Olsen reported clear and stable weather conditions in Illinois.

\begin{table*}[tbp]
  \setlength{\extrarowheight}{2.5pt}
  \centering
  \caption{Locations, setups and image acquisition parameters of successful observers, and disappearance / reappearance times derived from obtained light curves.}
  \label{tab:table2}
    \begin{tabular} { >{\raggedright}p{4.5cm} >{\centering}p{4.1cm} >{\centering}p{4.0cm} >{\centering\arraybackslash}p{4.0cm}}
    \hline\hline
    {Observer} & {Schindler \& Wolf} & {Olsen} & {Bardecker} \\
    \hline
    Closest city & Alder Springs, CA & Urbana, IL & Gardnerville, NV \\
    Latitude N (deg mm ss.ss) & 37 04 13.50 & 40 05 12.40 & 38 53 23.53 \\
    Longitude W (deg mm ss.ss) & 119 24 45.00 & 88 11 46.30 & 119 40 20.32 \\
    Altitude (m) & 1405 & 224 & 1534 \\
    Telescope type & Ritchey-Chrétien & Newtonian & Schmidt-Cassegrain\\
    Aperture, focal ratio & 600~mm, f/8 & 500~mm, f/4 & 304~mm, f/3.3 \\
    Camera & Andor iXon DU-888E-C00-BV & Watec 120N+ & MallinCam B/W Special \\
    Sensor type & Frame transfer EMCCD (\mbox{e2v~CCD201-20}, \mbox{back-illuminated}, grade~1) & Interline transfer CCD (Sony~ICX418ALL) & Interline transfer CCD (Sony~ICX428ALL-A with micro lenses) \\
    Sensor cooling (temperature\tablefootmark{a}) & Thermo-electrical ($-80\degree \mathrm{C}$) & None ($\approx -9\degree \mathrm{C}$) & None ($\approx +4\degree \mathrm{C}$) \\
    Frame grabber card & Andor CCI-24 PCIe & Pinnacle DVC-100 & Hauppauge USB-Live2 \\
    File format & FITS, uncompressed & Video, uncompressed & Video, \mbox{Lagarith\tablefootmark{b} codec}\\
    GPS time logger & Spectrum Instruments TM-4, triggered by electronic shutter of camera & Kiwi Video Time Inserter (VTI), overlaying GPS time stamp on camera frames & IOTA Video Time Inserter (VTI), overlaying GPS time stamp on camera frames \\
    Integration time (s) & 2.000 & 4.271 & 2.135 \\
    {Camera-internal frame accumulation} & none & 128~\texttimes~1/29.97 & 64~\texttimes~1/29.97 \\
    Cycle time per frame (s) & 2.00345 & 0.03337 & 0.03337 \\
    {Disappearance (D) time from square-well fit (UTC)} & 10:19:24.356~\textpm~0.159 & 10:18:05.447~\textpm~1.032 & 10:19:29.151~\textpm~0.452 \\
    {Reappearance (R) time from square-well fit (UTC)} & 10:19:50.249~\textpm~0.159 & 10:18:26.802~\textpm~1.032 & 10:19:48.370~\textpm~0.452 \\
    Duration of event (s) & 25.893~\textpm~0.225 & 21.355~\textpm~1.459 & 19.219~\textpm~0.639 \\ 
    Chord length (km) & 641.0~\textpm~3.9 & 525.9~\textpm~25.4 & 475.7~\textpm~11.2 \\
    SNR\tablefootmark{c}, star \& target combined & $\approx 12.6$ & $\approx 4.1$ & $\approx 4.7$ \\ 
   \hline    
\end{tabular}
\tablefoot{None of the observers have used any broadband filter. 
\tablefoottext{a}{Given temperatures are measured at the image sensor (Schindler \& Wolf) or close to the telescope setup at the time of recording (Olsen, Bardecker).}
\tablefoottext{b}{Lossless video compression algorithm.}
\tablefoottext{c}{Signal-to-noise (S/N) ratio per data point.}
}
\end{table*}

Schindler~\&~Wolf were observing with an Andor~iXon~DU\-888 camera with a back-illuminated EMCCD sensor in 2\texttimes2 binning mode. Being cooled to $-80\degree \mathrm{C}$ via an internal thermo-electrical cooler, exposures are virtually free of dark current. The frame-transfer design of the EMCCD allows a fast shift of the accumulated charge from the light-sensitive image area to a masked storage area that is read out subsequently, while the next exposure is already under way in the image area. This allows virtually gap-free imaging, with a dead time between frames of only 3.45~ms (the time to shift an image from the light-sensitive to the storage area). The video signal is digitized in the camera and transmitted to a proprietary PCIe frame grabber and controller card. Image acquisition was controlled with Andor SOLIS running on \mbox{Windows 7} \mbox{64 bit}. Running the camera in frame-transfer mode, subsequent series of 50~frames with 2~s integration time were taken and written directly to the PC's hard drive into three-dimensional FITS files without buffering image data in the computer's main memory (spooling). Between two image series, a gap of about 300~ms cannot be avoided owing to the non-real-time behavior of the operating system. A number of 50~images per series was chosen as a compromise between achieving a high probability to capture the disappearance and reappearance event during an image series while still achieving a timing accuracy well below 1~ms. Precise time stamps were logged with a Spectrum-Instruments TM-4 GPS receiver that was triggered directly via a TTL signal from the shutter port of the camera every time a series of 50 images started. Time stamps of the 2\textsuperscript{nd}~--~50\textsuperscript{th} image in a series have been extrapolated based on the time stamp of the first image and the well-known cycle time. The total uncertainty of the time stamp for each exposure is estimated to be \textless~0.25~ms and consists of the following components: 

   \begin{enumerate}[(a)]
      \item the uncertainty of the TM-4~GPS time measurement of \textpm~10~ns \citep{SpectrumInstruments2014},
      \item the synchronization of the TTL signal generated at the shutter port of the camera with the beginning of an exposure better than 1~\textmu s (Andor Technology, personal communication),
      \item cycle time that is known with an accuracy of \textpm~5~\textmu s, which leads to an accumulated timing uncertainty of \textpm~245 \textmu s after extrapolating the time stamp for the 50\textsuperscript{th} frame of a series. 
   \end{enumerate}

Consequently, the cycle time uncertainty dominates the total uncertainty of the time measurement. The event was captured in frames 30~--~43 of an image series. A time-lapse animation of the frames acquired during the event is provided in Figure~\ref{fig:animation1} and online. The setup was inspired by an earlier system described by \citet{Souza2006} and is based on very good experiences with this type of camera during characterization measurements of the SOFIA telescope \citep{Pfueller2012} and the subsequent upgrade of SOFIA's focal plane imager \citep{Wolf2014}. Differential aperture photometry was performed relative to five field stars in AstroImageJ\footnote{\url{http://www.astro.louisville.edu/software/astroimagej/}} \citep{Collins2013,Collins2015,Collins2016}, an image processing program written in Java on the basis of ImageJ \citep{Schneider2012}. The aperture radius was initially set to 3~pixels and subsequently varied based on the estimated full width at half maximum (FWHM) of the star image (radial profile mode). The background annulus size was set to an inner radius of 8 and an outer radius of 16~pixels; no background stars were visible within the annuli. 

\begin{figure}[tb]
   \centering
   \animategraphics[loop,autoplay]{5}{Animation01_frame_}{0001}{0025}
      \caption{Time-lapse animation of the stellar occultation as recorded by Schindler~\&~Wolf (click here to view the animation in a web browser). The time stamp indicates the start of each individual exposure; frames have been cropped to 210 x 210 pixels. }
   \label{fig:animation1}
\end{figure}

Olsen and Bardecker used monochrome video cameras that employ uncooled interline CCDs. Ambient temperature during data recording was $-9\degree\mathrm{C}$~(Olsen) and $+4\degree\mathrm{C}$~(Bardecker); sensor temperatures could not be measured and were somewhat higher than ambient owing to heat dissipated by the camera's sensor and electronics. An interline CCD has a masked storage column next to each light sensitive imaging column. All accumulated charges are simultaneously shifted sidewards into masked columns that are read out subsequently while the next frame is already acquired in the light sensitive area. Thanks to this design, these image sensors are also virtually free of any dead time and widely used in video cameras. The disadvantages compared to frame-transfer CCDs (that mostly find application in scientific cameras) are their reduced light-sensitive area and less optimized noise characteristics. The disadvantage of a reduced fill factor is partly mitigated by the use of micro lenses that are placed on each pixel of the CCD, which is the case in the MallinCam~B/W Special used by Bardecker. This results in a higher sensitivity compared to the Watec~120N+ used by Olsen that has a sensor without micro lenses. Both types of cameras are widely used by amateurs who routinely conduct observations of occultations and report their results to the International Occultation Timing Association (IOTA). Based on the monochrome EIA video signal standard \citep{EIA1957}, these cameras are running with a fixed frame rate of 29.97~frames/s (59.94~fields/s) and have an analog video output. One frame has 525~lines and consists of two fields that are interlaced (first field even-numbered lines, second field odd-numbered lines). To achieve longer integration times, a number of \emph{n} frames can be accumulated internally in the camera. Since the camera and its output are running at a fixed frame rate, the camera provides an accumulated frame \emph{n}-times at its output, while it is accumulating the next frame (see Figure~\ref{fig:light-curves}). A video time inserter (VTI) was used to overlay a GPS-provided time stamp on the lower part of each field coming from the camera's analog video signal output. The time-stamped analog video signal was then digitized directly to a Windows PC via a USB video capture device by using the software VirtualDub\footnote{\url{http://www.virtualdub.org/}}. Olsen  captured the video into uncompressed video files of 4~min length, while Bardecker used the lossless Lagarith video codec and recorded in segments of 5~min. The event was captured by both observers well within a video segment. 

Data reduction differs from classical CCD imaging in two ways: The camera returns the same accumulated frame \emph{n} times, superimposed with noise from analog amplifiers and the analog-digital conversion of the video capture device. This means that the accumulated frame is sampled \emph{n} times, so the average value of all samples in the respective bin is a representative value of the measured signal. Therefore, data have to be binned accordingly over \emph{n} frames, corresponding to each image accumulation interval. During data reduction, internal camera delays specific to each camera model (e.g.~the accumulated image is delayed by a number of fields at the camera output) and VTI (e.g.~the time stamp is delayed by a field) need to be considered to precisely establish the start of each individual exposure. 

The captured digital video has been analyzed with Tangra~3 \citep{Pavlov2014} and Limovie (K. Miyashita\footnote{\url{http://www005.upp.so-net.ne.jp/k_miyash/occ02/limovie_en.html}}) to extract the uncorrected time stamp for each frame from the video and to conduct differential aperture photometry of the target and field stars. While a classical annulus was used in Olsen's dataset for background estimation, two annulus sectors had to be used in Bardecker's dataset as a satellite trail was recorded in direct proximity of the target star in the frame integrated from 10:19:48.370 to 10:19:50.505 UTC. This accumulated frame recorded the reappearance of the target star at Bardecker's location; the profile of the target star could be clearly distinguished from the background. The two annular sectors were carefully positioned to avoid any pixels being contaminated by the satellite trail for local background estimation. A slight contamination of the central aperture could not be avoided, which resulted in a slightly overestimated flux of the target star, as seen in Figure~\ref{fig:light-curves} in the first data point after the estimated end of the occultation at Bardecker's site. Usage of a classical annulus for background estimation would have caused an overestimated background level, similar in counts to the measurement in the central aperture on the reappearing star, thus hiding the reappearance of the star in the derived light curve, and overestimating the length of the event at Bardecker's location by 2.135~s.

Time stamp corrections were applied as described in \citet{George2014} based on measurements by G.~Dangl\footnote{\url{http://www.dangl.at/ausruest/vid_tim/vid_tim1.htm}}, with the help of the software package R-OTE\footnote{\url{http://www.asteroidoccultation.com/observations/NA/}} \citep{George2013}: For Bardecker's dataset, a correction of $-2.1355~\mathrm{s}$ accounting for camera delay and $-0.01668~\mathrm{s}$ accounting for VTI delay was made; for Olsen's dataset, applied corrections were $-4.2876~\mathrm{s}$ and $-0.01668~\mathrm{s}$, respectively.

The difference in S/N between the three datasets is caused by differences in telescope aperture size, noise characteristics of the respective camera (e.g. dark noise, amplifier noise, read noise of A/D converters), and sensor design affecting quantum efficiency (back-illuminated frame-transfer CCD with fill factor~=~1 vs. front-illuminated interline CCD with fill factor~\textless~1, with or without micro lenses; differences in AR coating). Given the small telescopes used in this campaign, the low S/N was a necessary compromise to sample the faint star at an acceptable cadence, while integration times have been chosen wisely to go to the limit of the respective equipment.

\subsection{Far-infrared observations with \textit{Herschel}/PACS} \label{sec:firdata}

In addition to the data obtained from the occultation, we have revisited far-infrared photometric data that were previously acquired with the PACS photometer \citep{Poglitsch2010} on-board \textit{Herschel} \citep{Pilbratt2010}. Table~\ref{tab:table3} provides a summary of all observations of 2007~UK\textsubscript{126} that were made on 08~and~09 August 2010. This dataset has been reduced by constructing double-differential images optimized by source matching, which leads to a higher S/N  compared to supersky-subtracted images used in a previous reduction by \citet{Santos-Sanz2012}. Also, the PACS pipeline flux calibration is based on standard stars and asteroids that are significantly brighter than the target. The absolute photometric calibration of our reduction is instead based on a number of faint standard star measurements. The applied data reduction technique is described in detail in \citet{Kiss2014}. The fluxes obtained in each band are summarized in Table~\ref{tab:table4}.

\begin{table}[tbp]
  \setlength{\extrarowheight}{2.5pt}
  \centering
  \caption{Summary of observations of 2007~UK\textsubscript{126} with \textit{Herschel}/PACS.}
  \label{tab:table3}
    \begin{tabular}{cccc}
    \hline\hline
    OBSID & Start time (UTC) & Bands & $\beta$~(deg) \\
    \hline
    1342202277 & 2010 Aug 08 17:52:17 & B/R   & 110 \\
    1342202278 & 2010 Aug 08 18:02:48 & B/R   & 70 \\
    1342202279 & 2010 Aug 08 18:13:19 & G/R   & 110 \\
    1342202280 & 2010 Aug 08 18:23:50 & G/R   & 70 \\
    1342202324 & 2010 Aug 09 12:47:55 & B/R   & 70 \\
    1342202325 & 2010 Aug 09 12:58:26 & B/R   & 110 \\
    1342202326 & 2010 Aug 09 13:08:57 & G/R   & 110 \\
    1342202327 & 2010 Aug 09 13:19:28 & G/R   & 70 \\
    \hline
    \end{tabular}
\tablefoot{
OBSID denotes a \textit{Herschel} internal ID for each observation. All observations were done in standard mini scan-map mode with a scan-leg length of 3\arcmin, a scan-leg separation of 4\arcsec, and a total of 10 scan legs, resulting in a duration of 603~s per acquisition. For more details on the mini scan-map mode, its data reduction and calibration, see \citet{Balog2014}. The scan-maps were taken with the specified orientation angle $\beta$~relative to the instrument. During each observation, data were taken simultaneously in two bands: B - blue band (60~-~85~\textmu m), G - green band (85~-~120~\textmu m), R - red band (130~-~210~\textmu m). All observations were conducted with a repetition factor of 2, i.e. all maps have been observed twice.
}
\end{table}

\begin{table*}[tbp]
  \setlength{\extrarowheight}{2.5pt}
  \centering
  \caption{\textit{Herschel}/PACS fluxes of 2007~UK\textsubscript{126}, derived from a combination of all data available in a given band.}
  \label{tab:table4}
    \begin{tabular}{ccccccccc}
    \hline\hline
          &       &       & \multicolumn{3}{c}{\textit{Herschel}-centric} &       &       &  \\
    Band  & $\lambda_\mathrm{ref}$ (\textmu m) & Mid-time & $r_\mathrm{H}$ (AU) & $\Delta$~(AU) & $\alpha$~(deg) & In-band flux (mJy) & FD (mJy) & $1\sigma$~(mJy) \\

\hline
    B     & 70    & 2010 Aug 09 03:30 & 44.94437 & 45.16375 & 1.27 & 11.43~\textpm~1.20 & 11.55 & 1.34 \\
    G     & 100   & 2010 Aug 09 03:51 & 44.94436 & 45.16351 & 1.27 & 13.20~\textpm~1.47 & 13.33 & 1.63 \\
    R     & 160   & 2010 Aug 09 03:41 & 44.94437 & 45.16363 & 1.27 & 8.08~\textpm~2.03 & 8 & 2.05 \\
    \end{tabular}
\tablefoot{
For all details on the data reduction, see \citet{Kiss2014}. The photometric uncertainty is derived from aperture photometry on 200 artificial sources, resembling the PACS PSF in the respective band, that have been planted individually on the double-differential image. The distribution of measured artificial source fluxes resembles a Gaussian, so its standard deviation defines the photometric uncertainty of the respective image. $\lambda_\mathrm{ref}$ - reference wavelength for respective band, $r_\mathrm{H}$ - heliocentric distance at mid-time, $\Delta$ - \textit{Herschel}-centric distance at mid-time, $\alpha$~- phase angle at mid-time, FD - color-corrected monochromatic flux density at reference wavelength, $1\sigma$~- estimated uncertainty of FD (includes the 5\% absolute flux calibration error of PACS).
}
\end{table*}

We have also checked data from the \textit{Wide-field Infrared Survey Explorer} \citep[{\it WISE},][]{Wright2010} satellite for possible detections of 2007~UK\textsubscript{126}. Although two possible detections in images taken on 10 January 2010 are listed in the \textit{WISE} All-Sky Known Solar System Object Possible Association List\footnote{\url{http://irsa.ipac.caltech.edu/cgi-bin/Gator/nph-scan?submit=Select&projshort=WISE}}, we concluded that those are false positives for several reasons: The associated sources are off by several arcsec from their predicted positions, and our thermophysical modeling (see Section~\ref{sec:tpm}) predicts fluxes in the \textit{WISE} bands that are well below the respective detection limits. Both \textit{Herschel}/PACS and \textit{WISE} measurements were taken close in time, i.e. at very similar observing geometry, so it appears safe to say that 2007~UK\textsubscript{126} was below the detection limits of \textit{WISE}. Also, the Minor Planet Center does not list any entries for 2007~UK\textsubscript{126} reported from \textit{WISE}.


\section{Results from the stellar occultation}
\subsection{Occultation light curves and square-well fits} \label{sec:squarewellfits}

The integration time and achieved S/N of each camera dictate the fundamental limit of how accurate the moment of disappearance and reappearance of the occulted star can be determined on each chord. The expected light curve resembles a square-well if the following conditions are fulfilled:
\begin{enumerate}
\item{The body has no atmosphere, or one that is so thin that it cannot be detected, given the sampling rate and S/N.}
\item{Effects of Fresnel diffraction and the finite stellar diameter are negligible.}
\end{enumerate}

Considering the peak sensitivity wavelength of the Si-based CCD cameras ($\lambda \approx 650~\mathrm{nm}$) and the geocentric distance of 2007~UK\textsubscript{126}, the Fresnel scale is $F \approx 1.44~\mathrm{km}$ \citep{Roques1987}. The estimated angular stellar diameter is about $0\farcs0097$ \citep{vanBelle1999}, or 0.3~km projected at the geocentric distance of 2007~UK\textsubscript{126}. Given the very high speed of the shadow on the Earth's surface of about $v = 24.1~\mathrm{km~s^{-1}}$ and the three orders of magnitude larger size of 2007~UK\textsubscript{126}, it becomes clear that both effects can be neglected for the dataset at hand.

\begin{figure*}[htb]
   \centering
   \includegraphics{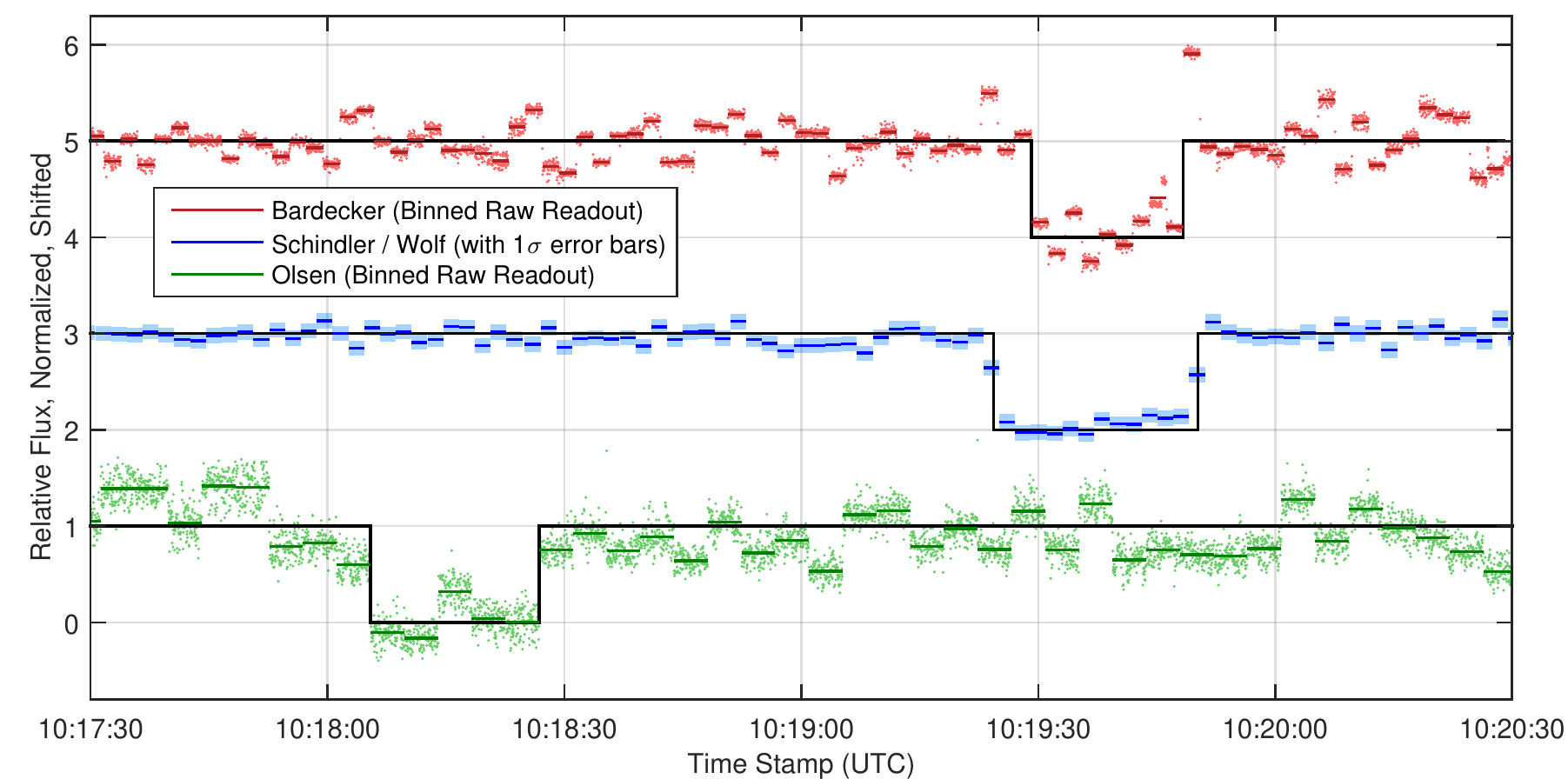}
   \caption{Obtained light curves for the occultation on 15~November~2014. The occultation was recorded at different times due to different longitudes and latitudes of the observatories; the shadow was moving from east to west (c.f.~Figure~\ref{fig:shadow-path}). Light curves were normalized based on the combined relative flux of the star \mbox{UCAC4 448-006503} and TNO {2007~UK\textsubscript{126}}, averaged over all available images before and after the event. The occulted star's apparent V-band magnitude is $m_\mathrm{V} = 15.8~\mathrm{mag}$, which was four magnitudes brighter than {2007~UK\textsubscript{126}} ($m_\mathrm{V} \approx 19.8~\mathrm{mag}$). Raw data from Bardecker and Olsen is plotted in light red and light green; their video cameras were running at a fixed frame rate, accumulating \emph{n} frames internally and returning the result $n$ times at the analog video output  (see Section~\ref{sec:observations} for details). Consequently, the relative flux was averaged over each bin of 64 (Bardecker, dark red) and 128 samples (Olsen, dark green; see discussion in text). Data points by Schindler~\&~Wolf (blue) were derived from single images, with relative flux errors ($1\sigma$) plotted in light blue. The black lines represent square-well fits that were used to derive the disappearance and reappearance times for each chord (see text for details).}
   \label{fig:light-curves}
    \end{figure*}

The light curves that have been obtained by the three observers are illustrated in Figure~\ref{fig:light-curves}, together with derived square-well fits. All relative fluxes have been normalized based on the average combined relative flux of star and TNO in all available exposures before and after the occultation (``baseline''). None of the light curves indicate a gradual decline and emergence of the star, so a square-well fit is a legitimate approximation. Given the achieved sampling rates, the presence of a thin atmosphere cannot be excluded; this question remains to be studied during future occultations that allow for faster sampling (i.e. brighter star or availability of larger telescopes). Given the quasi-zero dead time of all cameras used in this study, disappearance and reappearance of the occulted star must have occurred (with near-100\% certainty) during integration of an exposure (a single frame for Schindler/Wolf, or \emph{n} accumulated frames for Bardecker and Olsen). The measured relative flux in the respective exposure is consequently smaller than the combined relative flux of the star and the TNO, but larger than the TNO's relative flux alone.

The S/N of the dataset of Schindler~\&~Wolf is sufficient to clearly isolate the frames that recorded disappearance and reappearance, and to interpolate their time stamps to subframe accuracy. As the response of the CCD is linear, the ratio between the measured relative flux in the disappearance or reappearance frame and the baseline is directly proportional to the offset of disappearance or reappearance in time relative to the start of the respective exposure\footnote{This simplification is only accurate for a body without any atmosphere. Assuming a gradual dimming would have happened on shorter time scales than the integration time, the gradual transition could have been integrated by a single image. The length of the chord would then be slightly overestimated.}. An upper and lower limit of this ratio has been derived using the normalized relative flux error $\sigma_\mathrm{Signal} = {\left(\mathrm{S/N}\right)^{-1}}$, which propagates directly into the uncertainty of the estimated disappearance and reappearance times $\sigma_\mathrm{D/R} = \sigma_\mathrm{Signal}~t_\mathrm{int}$, as given in Table~\ref{tab:table2}.

Unfortunately, the S/Ns of the datasets of Olsen and Bardecker do not allow for the identification of the accumulated frame that recorded disappearance and reappearance. This becomes clear from the light curves: None of the accumulated frames has a relative flux that separates it from the upper or lower baseline with statistical significance. Thus, we decided to apply square-well fits that are coincident with the sampling frequency. Again, the normalized relative flux error has been used as an uncertainty estimate to calculate timing uncertainties (see Table~\ref{tab:table2}). We note that the R-OTE software package uses the Akaike information criterion (AIC) as a logic to decide objectively if a square-well fit shall be applied to exposure timing (3 parameter model) or sub-exposure timing (4 parameter model). While the goodness of fit improves with an additional parameter, the fit is not necessarily a better representation of the data at hand. By weighing the number of parameters of a model against its goodness of fit, AIC provides a formal way to decide if sub-exposure timing can be applied. For the two datasets of Olsen and Bardecker, the AIC ruled in favor of the simpler model.

\subsection{Size of the shadow from a best-fitting ellipse}

The reconstruction of shadow size and shape is done in the geocentric fundamental (Besselian) plane, which is perpendicular to the line connecting the apparent position of the occulted star UCAC4~448-006503 (at the epoch of the observation, transformed from the J2000 UCAC4 position) and the position of {2007~UK\textsubscript{126}} as given by the ephemeris. In the fundamental plane, the shadow of {2007~UK\textsubscript{126}} is not distorted by Earth’s curvature, i.e. shape and size of the shadow can be directly derived. Each observer’s position on the Earth’s surface (latitude, longitude) is projected twice onto the fundamental plane:  At the disappearance (D) and reappearance (R) time of the star as measured by the respective observer. This results in six event coordinates on the fundamental plane that represent points on the limb profile of {2007~UK\textsubscript{126}}. A reference frame is then defined on the fundamental plane, which is fixed with respect to the shadow of {2007~UK\textsubscript{126}}, or in other words, moving in the fundamental plane with the shadow. After transforming the D and R coordinates from the fundamental plane into the moving reference frame, the shadow is reconstructed. Details of this reduction approach can be found in, for example, \citet{Millis1989}, \citet{Millis1979} and \citet{Wasserman1979}.

Based on the assumption that 2007~UK\textsubscript{126} is of sufficient size that its shape can be approximated either by a Maclaurin or a Jacobi ellipsoid, its projected shadow on the Earth's surface is expected to be an ellipse. For a subsequent analysis of size and shape, we need to find the ellipse that best fits to the measured disappearance (D) and reappearance (R) times (or respectively, the D and R coordinates in the reference frame moving with the shadow). An ellipse can be described by five geometrical parameters: Its center coordinates $x_\mathrm{El}, y_\mathrm{El}$, the major and minor axes $a_\mathrm{El}$ and $b_\mathrm{El}$, and the orientation angle of the major axis $\theta_\mathrm{El}$. Fortunately, the availability of three chords covering both sides of the shadow path allows for an ellipse fit without taking additional assumptions.

We used Occult\footnote{\url{http://www.lunar-occultations.com/iota/occult4.htm}}, a software program written and continuously developed by D. Herald and used by occultation observers worldwide for predictions of events and subsequent data analysis. Using Occult, we translated measured D and R times and their derived uncertainties (see previous section) into D and R coordinates with propagated uncertainties in the moving reference frame, and calculated an initial ellipse fit \citep[for details, see][]{OccultHelp}. To verify these results independently, we recalculated the ellipse fit with our own code written in Matlab and realized that the algorithm implemented in Occult has a significant shortcoming. Occult uses a direct least-square ellipse fit algorithm, where uncertainties of D and R timings have no effect on the resulting ellipse and are not propagated into the uncertainties associated to the ellipse parameters. Timing error estimates are collected merely for archiving purposes at this moment. The uncertainties of the geometric ellipse parameters provided by Occult are solely derived from the residuals between the ellipse and each D or R coordinate, measured along the line between ellipse center and D or R. Although Occult allows the user to specify weights to differentiate between different levels of data quality, these weights only influence the estimated uncertainties of the ellipse parameters, not the ellipse fit itself.

It became clear that the desired propagation of timing errors into the ellipse fit leads to a non-linear errors-in-variable problem. We found a perfectly suited algorithm recently described by \citet{Szpak2015} that offers a solution for exactly this type of problem. The D and R uncertainties can be described as independent Gaussian noise with zero mean that is inhomogeneous, i.e. that depends on the respective setup, integration time and observing conditions. Therefore, we can use the propagated D and R uncertainties in the moving reference frame to define a covariance matrix for each D and R {coordinate}. The algorithm uses these covariance matrices as weights in a cost function that needs to be minimized to determine the best-fitting ellipse. After the algorithm has calculated the algebraic parameters of the ellipse and their corresponding covariances, a conversion and covariance propagation to the ellipse's geometrically meaningful parameters is conducted. In this way, we get a reliable and robust estimate of the uncertainties of all parameters of interest, while the ellipse fit itself considers the noise in the data points. Finally, a confidence region in the plane of the ellipse is calculated, illustrating the zone in which the true ellipse is located at a given probability. We chose a planar 68.3\% confidence region to be consistent with other literature in the field instead of the 95\% region suggested by \citet{Szpak2015}, e.g. for industrial machine vision applications.

\begin{figure}[tb]
   \centering
   \includegraphics{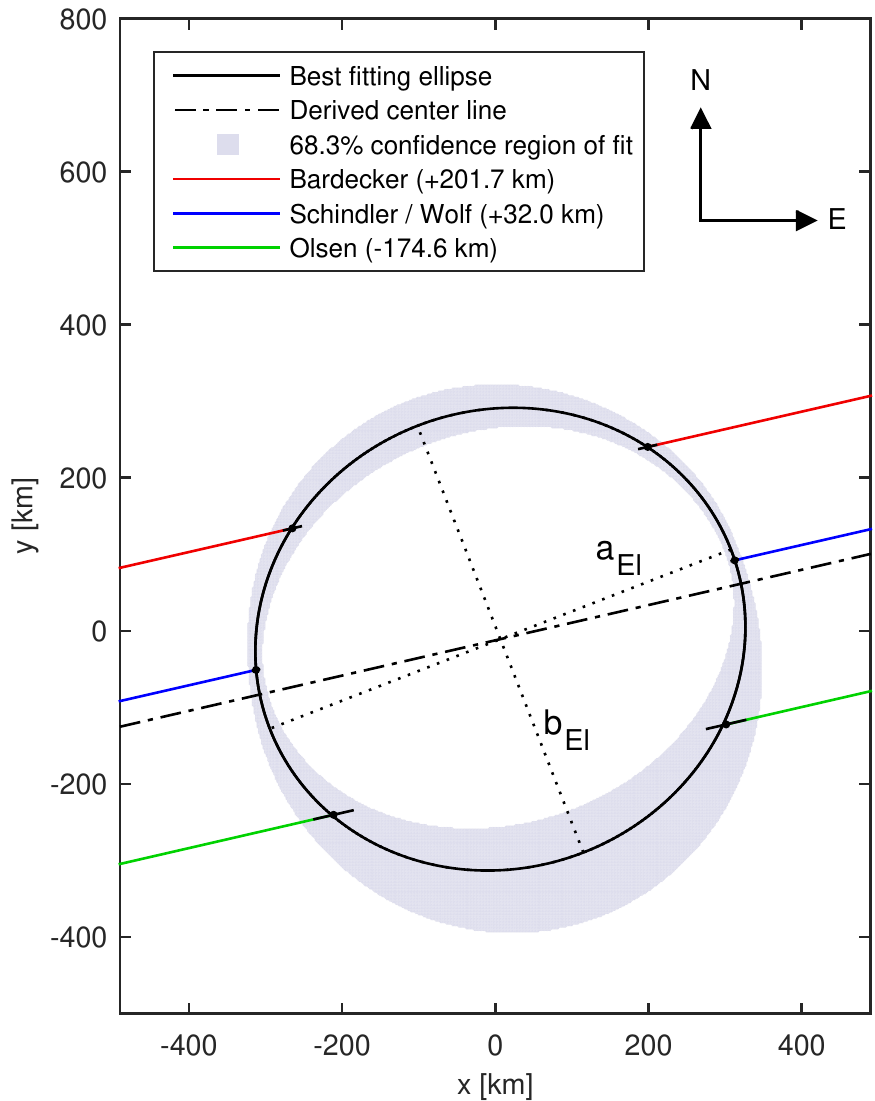}
   \caption{The derived ellipse fit and its 68.3\% confidence region (gray). Uncertainties of each individual ellipse parameter are summarized in Table~\ref{tab:table5}. The angle $\theta$ gives the rotation of the major axes, measured from east to north. The plotted error bars of the D and R locations have been derived from the timing deviations given in Table~\ref{tab:table2}. The major and minor axes are indicated by the dotted lines. The distances provided in the legend are topocentric with respect to the centerline derived for the best fit (dash-dotted line). The chord by Schindler \& Wolf sampled 2007~UK\textsubscript{126} almost at the {centerline}, while the chords by Bardecker and Olsen sampled the upper and lower part quasi symmetrically. The geocentric distance of 2007~UK\textsubscript{126} was $r_\mathrm{G} = 42.572~\mathrm{AU}$ at the time of the observation.}
   \label{fig:ellipse-fit}
    \end{figure}

The resulting geometrical parameters of the best fitting ellipse and their associated uncertainties are listed in Table~\ref{tab:table5}. The ellipse fit and its 68.3\% confidence region is illustrated in \mbox{Figure~\ref{fig:ellipse-fit}}. The positive y axis is directed to the north, while the positive x axis indicates east. The origin (x,y) = (0,0) represents the average of the minimum and maximum event coordinate in north-south and east-west direction.

\begin{table}[tbp]
  \setlength{\extrarowheight}{2.5pt}
  \centering
  \caption{Geometric parameters of the best fitting ellipse and associated uncertainties.}
  \label{tab:table5}
    \begin{tabular}{cc}
    \hline\hline
    Parameter & Value \\
    \hline
    $a_\mathrm{El}$ (km) & $645.80 \pm 5.68$ \\
    $b_\mathrm{El}$ (km) & $597.81 \pm 12.74$ \\
    $x_\mathrm{El}$ (km) & $6.57 \pm 2.06$ \\
    $y_\mathrm{El}$ (km) & $-10.82 \pm 3.13$ \\
    $\theta_\mathrm{El}$ (deg) & $21.25 \pm 5.65$ \\
    \hline
    \end{tabular}
\end{table}

\begin{figure}[tb]
   \centering
   \includegraphics{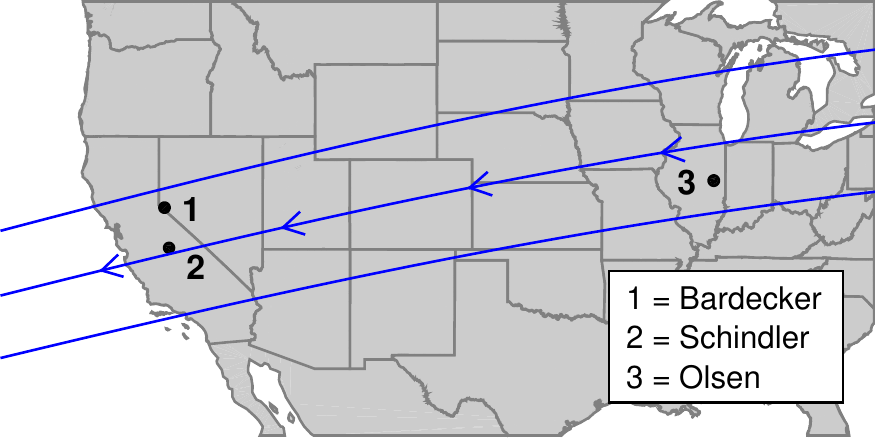}
   \caption{Reconstructed shadow path and geographic locations of all observers across the continental US. The three lines indicate the shadow path's northern boundary, centerline, and southern boundary, assuming an elliptical shape for 2007~UK\textsubscript{126} based on the ellipse fit presented in Figure~\ref{fig:ellipse-fit}. The arrows indicate the direction of travel of the shadow on the Earth's surface from east to west. An animation illustrating the event and the acquired light curves is provided as supplementary material on the A\&A website.}
   \label{fig:shadow-path}
    \end{figure}

Figure~\ref{fig:shadow-path} illustrates the reconstructed shadow path on the surface of the Earth, where the shadow's upper and lower boundary and centerline are based on the size and orientation of the ellipse as illustrated in Figure~\ref{fig:ellipse-fit}. An animation that illustrates the shadow traveling across the continental US and the measurements conducted in parallel is provided as supplementary material on the A\&A website. We conclude that the relative measurement uncertainty of the ellipse axes is in the order of 0.9\% and 2.1\%, respectively. The axial ratio of the ellipse translates to ${a_\mathrm{El}}/{b_\mathrm{El}} = 1.080 \pm 0.025$, i.e. a circular fit and hence a pole-on viewing geometry can already be ruled out geometrically.

\subsection{Albedo}
Thanks to the occultation measurement, we can improve the geometric albedo estimate. Following the formula
\begin{equation}
p_\mathrm{V} = 4 \cdot 10^{0.4 (H_\mathrm{V,Sun}-H_\mathrm{V,TNO})} \cdot \left(1\,\mathrm{AU}\right)^2~a_\mathrm{El}^{-1}~b_\mathrm{El}^{-1}
\label{eq1}
,\end{equation} 

where $H_\mathrm{V,Sun} = -26.78~\mathrm{mag}$ is the absolute magnitude of the Sun (Bessel V\footnote{Willmer (2006), \url{http://mips.as.arizona.edu/~cnaw/sun.html} using \citet{Bohlin2004} and \citet{Fukugita1995}\label{ftn:UoA}}), $H_\mathrm{V,TNO} = 3.69 \pm 0.044~\mathrm{mag}$ is the absolute magnitude of 2007~UK\textsubscript{126} in V-Band (see Section~\ref{sec:properties}, although this value might include a small flux contribution from a potential satellite) and $1\,\mathrm{AU} = 1.49598 \cdot 10^8~\mathrm{km}$. Considering propagation of all measurement errors (see Appendix~\ref{app:errors}), we derive $p_\mathrm{V} = 15.0 \pm 1.6 \%$. This calculation assumes that, between the epochs of the photometric measurements (September 2008) and the occultation (November 2014), the projected area did not change, i.e. the pole orientation of the body as seen from Earth remains virtually the same. This assumption is reasonable: As pointed out by \citet{Sheppard2002}, with reference to \citet{Harris1994}, an object of the size of 2007~UK\textsubscript{126} has a damping timescale for its non-principal axis rotation (``wobble'') that is significantly smaller than the age of the solar system. Considering the orbital period of 636.73 yr, a body in pure principal axis rotation would appear in almost the same projection on such short timescales. The new albedo estimate is within the previously derived range by \citet{Santos-Sanz2012} of $p_\mathrm{V} = 16.7^{+5.8}_{-3.8}\%$. 

2007~UK\textsubscript{126} is one of eight detached objects that have been observed with \textit{Herschel} in the far-infrared and subsequently analyzed via thermophysical modeling. Based on this sample, \citet{Lacerda2014} derived a median albedo for detached objects of $p_{\mathrm{V,DO}} = 16.7\%$. Considering our new estimate, the median albedo for this class of objects shifts to $p_{\mathrm{V,DO}} = 15.0\%$, but the sample size is too small for reliable statistical analysis.

Substituting the magnitudes in equation \ref{eq1} with the R-band estimates $H_\mathrm{R,TNO} = 3.07 \pm 0.044~\mathrm{mag}$ (see Section~\ref{sec:properties}) and $H_\mathrm{R,Sun} = -27.12~\mathrm{mag}$ (Bessel R\footref{ftn:UoA}), we obtain a geometric albedo in the R-band of $p_{R} = 19.5 \pm 2.0 \%$.

\subsection{Shape and size}
\label{sec:shape}

The very small light curve variations ($\Delta m_\mathrm{V} = 0.03 \pm 0.01~\mathrm{mag}$ peak-to-peak, see Section~\ref{sec:properties}) are another strong indicator of a Maclaurin spheroid with albedo variations that is not seen pole-on. From spacecraft fly-bys, we know about objects in the solar system that were shaped into oblate spheroids by their own gravity, but which are considerably smaller than 2007~UK\textsubscript{126}: The icy Saturnian satellites Mimas \citep[$\rho = 1.149 \pm 0.007~\mathrm{g~cm^{-3}}$, $d_\mathrm{eq} = 396.4 \pm 0.8~\mathrm{km}$,][]{Roatsch2009}, Enceladus \citep[$\rho=1.609 \pm 0.005~\mathrm{g~cm^{-3}}$, $d_\mathrm{eq} = 504.2~\mathrm{km}$,][]{Roatsch2009} and Miranda \citep[$\rho=1.20 \pm 0.14~\mathrm{g~cm^{-3}}$, $d_\mathrm{eq} = 472 \pm 3~\mathrm{km}$,][]{Jacobson1992}. We feel it is a reasonable assumption to constrain our following analysis to a Maclaurin spheroid that rotates around its principal axis $c$, and to discard the Jacobi solution.

Viewing a Maclaurin spheroid pole-on ($\theta = 0\degree$) would lead to a circular shadow during an occultation, while viewing it equator-on ($\theta = 90\degree$) would lead directly to the elliptical shadow that was measured. While a circular fit (and hence a pole-on view) has been ruled out geometrically, the ellipse derived in the previous section represents a distorted projection of the true shape of 2007~UK\textsubscript{126} in any other pole orientation than equator-on. For a Maclaurin spheroid, we can calculate the flattening ratio ${a}/{c}$ from the major axis $a_\mathrm{El}$ and minor axis $b_\mathrm{El}$ of the shadow ellipse and the angle $\theta$ between the spheroid's rotation axis and the line of sight:

\begin{equation} 
\frac{a}{c} = \sqrt{\frac{\sin^2 \theta}{\frac{b_\mathrm{El}^2}{a_\mathrm{El}^2} - \cos^2 \theta}}
\label{eq2}
.\end{equation}

\begin{figure}[tb]
   \centering
   \includegraphics{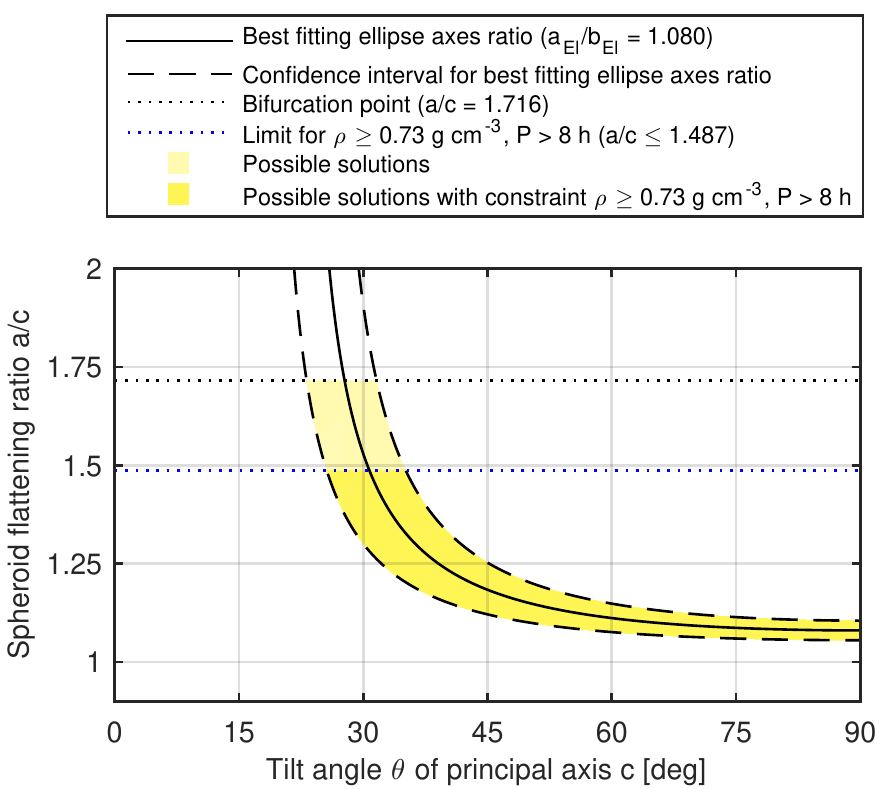}
   \caption{Flattening ratio $a/c$ of the Maclaurin spheroid as a function of angle $\theta$ between the spheroid's rotation axis and the line of sight. Without a density estimate, true flattening ratios up to the bifurcation point are theoretically plausible. We argue that a more realistic lower density limit is $\rho = 0.73~\mathrm{g~cm^{-3}}$, which would exclude extreme axis ratios and therefore constrain the volume of the body considerably.}
   \label{fig:flattening-ratio}
    \end{figure}

We do not know $\theta$ at the time of the occultation. This implies that we can only specify a plausible range of flattening ratios based on the uncertainties of the shadow ellipse parameters and an arbitrary tilt angle. The upper limit of the flattening ratio is given by the bifurcation point between a Maclaurin spheroid and a Jacobi ellipsoid at ${a}/{c} = 1.71609$, corresponding to an ellipse eccentricity of $e = 0.81267$ \citep{Chandrasekhar1967}. As illustrated in Figure~\ref{fig:flattening-ratio} and summarized in Table~\ref{tab:table6}, the flattening ratio of 2007~UK\textsubscript{126}, and therefore its size, are relatively poorly constrained.
 
We also do not have a density estimate for 2007~UK\textsubscript{126}. Models of volatile retention \citep[see e.g.][]{Schaller2007} could be used to derive a lower bulk density limit when surface ices or a thin atmosphere are present, but the featureless near-infrared spectrum (see Section~\ref{sec:properties}) does not indicate the presence of any prominent volatiles (within the S/N limits of the dataset). As pointed out by \citet{Brown2013}, the absence of volatiles and a measurable atmosphere cannot be used as an argument to derive an upper density limit, as Jeans escape is unfortunately not the only mechanism that can cause a body to lose its volatiles, even though it is the slowest. 

One way to estimate a lower bulk density limit would be a precise determination of the rotation period. At bifurcation, we can calculate the density of the spheroid as shown by \citet{Chandrasekhar1967} from 
\begin{equation}
\rho = \frac{\omega^{2}}{0.37423~\pi~G}
,\end{equation} 
where $\omega$ denotes the angular velocity and $G = 6.67384 \cdot 10^{-11}~\mathrm{m^{3}~kg^{-1}~s^{-2}}$ the gravitational constant. We only know that 2007~UK\textsubscript{126} takes more than $P > 8~\mathrm{h}$ for a full rotation (see Section~\ref{sec:properties}), which means its bulk density at bifurcation would be $\rho < 0.61~\mathrm{g~cm^{-3}}$. Rotation period and density are indirectly proportional to the square, i.e. if the body takes twice as long for a full rotation, its lower density limit would be one quarter of this estimate. We can therefore not constrain a lower limit on bulk density for 2007~UK\textsubscript{126} at this time. 

Estimated bulk densities and sizes of TNOs have been collected, e.g. in \citet{Ortiz2012} (see supplementary information), \citet{Brown2013b} and \citet{Johnston2014}. We compiled a list of all TNOs whose size has been constrained by thermophysical modeling and/or occultations, and bulk density has been estimated owing to the presence of a satellite with a known orbit (see Appendix~\ref{app:densities}). We did not include TNOs for which bulk density estimates have been derived solely from light curves (as an assumption on body figure has to be made in these cases), or limits on bulk density have been stated mistakenly based on the absence of volatiles or a measurable atmosphere. Our list covers 19 objects with a diameter range of $157 - 2374~\mathrm{km}$. We emphasize that this compilation can only be used for qualitative statements; it is not meaningful to derive a correlation function between size and density -- the sample is too small, could be observationally biased, and uncertainties are, in general, very large. Also, the large diversity among the TNO population is not understood, and bodies could have undergone entirely different evolutions. To our knowledge, no density estimate has been derived for any detached object to date. Only few objects have been studied through far-infrared observations, and 2007~UK\textsubscript{126} is the first detached object that has been studied in detail during an occultation. In addition, the limits of orbital elements of the detached object population are unclear, and transitional objects between the scattered disk and the inner Oort cloud could belong to this population as well. The only two scattered disk objects (SDOs) for which densities have been estimated thanks to their satellites are Ceto \citep[$d_\mathrm{{system,eff}} = 281 \pm 11~\mathrm{km}$, $\rho = 0.64^{+0.16}_{-0.13}~\mathrm{g~cm^{-3}}, p_\mathrm{V} = 5.6 \pm 0.6$,][]{Santos-Sanz2012} and Eris \citep[$d_\mathrm{eff} = 2326 \pm 12~\mathrm{km}$, $\rho = 2.52 \pm 0.05~\mathrm{g~cm^{-3}}, p_\mathrm{V} = 96^{+9}_{-4}\%$,][]{Sicardy2011}. Both objects could not be more contrary in character; their properties are at complete opposite ends of the parameter space, illustrating the difficulties at hand. 

From our list, we find that, except for two targets (both about half the size of 2007~UK\textsubscript{126}), no TNO has an estimated density $\rho < 0.6~\mathrm{g~cm^{-3}}$. This corresponds to our estimated lower density limit at bifurcation based on $P = 8~\mathrm{h}$. A subset of 13 objects in our list have a size $d_\mathrm{eff} < 800~\mathrm{km}$; the average density of this subset is $\bar{\rho} = 0.87~\mathrm{g~cm^{-3}}$, while the median density is $\rho_\mathrm{median} = 0.73~\mathrm{g~cm^{-3}}$. Eleven of these 13 objects are smaller than $d_\mathrm{eff} < 400~\mathrm{km}$, so this selection is strongly biased towards objects that are significantly smaller than 2007~UK\textsubscript{126}. 

We feel that it is unlikely that 2007~UK\textsubscript{126} has a bulk density below $\rho = 0.73~\mathrm{g~cm^{-3}}$ given its size, since this would require a significant porosity comparable to a comet. The Rosetta mission revealed a density of $\rho = 0.533 \pm 0.006~\mathrm{g~cm^{-3}}$ and a porosity of $72-74\%$ for the nucleus of 67P/Churyumov-Gerasimenko \citep{Paetzold2016}. These properties appear feasible for highly fractured rubble piles, but are hard to imagine for a dwarf planet candidate that is larger than the three Saturnian satellites mentioned earlier. Arguing from a different perspective,  finding an oblate spheroid with a flattening ratio close to bifurcation is very unlikely, since 2007~UK\textsubscript{126} is certainly not composed of a strengthless fluid. Solving the following equation provided by \citet{Chandrasekhar1967} numerically, 
\begin{equation} 
\rho = \omega^2 \left( \pi~G~\frac{\sqrt{1-e^2}}{e^{3}}~2~\left(3-2e^{2}\right) \arcsin e - \frac{6}{e^2} \left(1-e^2\right) \right)^{-1}
\label{eq3}
,\end{equation} 
leads to an ellipse eccentricity of $e = 0.7401$, or ${a}/{c} = 1.4870$, assuming $\rho = 0.73~\mathrm{g~cm^{-3}}$ and $P = 8~\mathrm{h}$. A longer rotation period would lower the flattening ratio. We feel that a flattening ratio of ${a}/{c} = 1.4870$ represents an acceptable, albeit qualitative, upper limit.

Table~\ref{tab:table6} summarizes the possible size range of all three spheroid axes for both the generic case (unknown density) and considering the added qualitative constraint on density. 
It can be seen that the occultation data is able to improve the previous size estimate of $d_\mathrm{Sphere,eff} = 599 \pm 77~\mathrm{km}$ for an equivalent sphere that was derived by \citet{Santos-Sanz2012} solely through thermophysical modeling based on \textit{Herschel}/PACS data. In the next section, we refine our size estimate from the occultation further through thermophysical modeling, using the occultation data as constraints.

\begin{table}[tbp]
  \setlength{\extrarowheight}{2.5pt}
  \centering
  \caption{Possible size range of the axes of a Maclaurin spheroid.}
  \label{tab:table6}
    \begin{tabular}{cc|c}
    \hline\hline
   Parameter  & $a/c < 1.7161$ & $a/c \leq 1.4870$\\
     &  & $P \geq 8~\mathrm{h}$\\ 
     & & $\rho \geq 0.73~\mathrm{g~cm^{-3}}$ \\
\hline
       $a = b$ (km) & $640-651$ & $640-651$ \\
       $c$ (km) & $373 - 611$ & $432 - 611$ \\
       $d_\mathrm{Sphere,eff}$ (km) & $535 - 638$ & $563 - 638$ \\
    \hline
    \end{tabular}
\tablefoot{The equivalent diameter of a sphere $d_\mathrm{Sphere}$ has been calculated to compare with a previous size estimate.}
\end{table}


\section{Results from thermophysical modeling}\label{sec:tpm}
To improve our geometrically obtained size estimate of 2007~UK\textsubscript{126}, we model the body's thermal emission using a thermophysical model (TPM) code. We conduct a parametric study to predict far-infrared fluxes for a wide range of plausible geometries and physical properties, and compare them to our re-reduced \textit{Herschel}/PACS measurements presented in Section~\ref{sec:firdata}. A description of the thermophysical model, its parameters and further details can be found in \citet{Mueller1998,Mueller2002}. The model considers the heliocentric distance of 2007~UK\textsubscript{126} ($r_\mathrm{H} = 44.944~\mathrm{AU}$) at the epoch of \textit{Herschel's} observations. 

\begin{figure}[tb]
   \centering
   \includegraphics{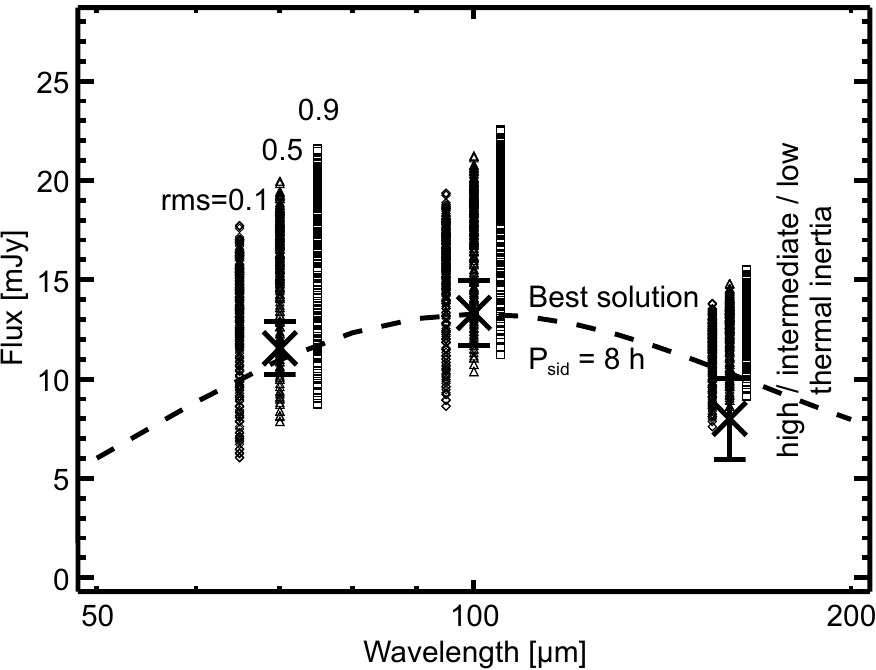}
   \caption{Far-infrared fluxes of 2007~UK\textsubscript{126} with estimated $1\sigma$ uncertainties as measured by \textit{Herschel}/PACS in three bands (see Table~\ref{tab:table4}), in comparison to far-infrared fluxes calculated from the modeled thermal emission of all cases covered by our parametric study, discarding a flux contribution from the satellite and assuming $P = 8~\mathrm{h}$. For clarity, flux predictions have been plotted with a slight offset in wavelength based on the applied surface roughness ($s = \{0.1, 0.5, 0.9\}~\mathrm{rms}$). In general, predicted fluxes increase with decreasing thermal inertia. The dashed line represents the best model obtained for an intermediate roughness ($s = 0.5~\mathrm{rms}$) with $\varGamma = 3~\mathrm{J~m^{-2}~s^{-0.5K^{-1}}}$, $\theta=70 \degree$ and $a/c=1.11$, resulting in $D_\mathrm{eff} = 618~\mathrm{km}$. See text for details.}
   \label{fig:FIR-fluxes}
    \end{figure}

2007~UK\textsubscript{126} is considered  a binary system. Since we do not know the position of the satellite relative to the primary during \textit{Herschel's} observations, we need to consider that the satellite might have contributed some flux to the \textit{Herschel}/PACS measurements. We therefore studied the following cases to constrain which combination of parameters are compatible to the occultation and to the \textit{Herschel}/PACS data:

\begin{enumerate}[I]
\item{The satellite did not contribute flux to the observed FIR fluxes;}
\item{The satellite contributed a maximum plausible flux to the observed FIR fluxes;}
\end{enumerate}
each considering 
\begin{enumerate}[(a)]
\item{A thermal inertia of \newline $\varGamma = \{0.3, 0.7, 1, 3, 5, 10\}~\mathrm{J~m^{-2}~s^{-0.5}~K^{-1}}$};
\item{A surface roughness with a slope of $s = \{0.1, 0.5, 0.9\}~\mathrm{rms}$};
\item{A rotation period of $P=\{8, 11.05, 14.30, 20.25\}~\mathrm{h}$}
\end{enumerate}
for three representative shadow ellipses: The best fitting ellipse, and the largest and smallest ellipse according to the estimated uncertainties of the major and minor axes. For each ellipse geometry, we considered the possible tilt angle range of the body's principal axis $c$ starting at the lowest limit constrained by bifurcation, and then discretized in multiples of $5\degree$ up to $\theta = 45\degree$, and in multiples of $10\degree$ between $\theta = 50\degree$ and an equator-on viewing geometry ($\theta = 90\degree$). The qualitatively derived limit of ${a}/{c} = 1.4870$ discussed in the previous section was discarded to enable an unbiased analysis. To study the described parameter space, 4320~flux predictions had to be made, not counting additional predictions to test, for example, an even more extreme thermal inertia. Figure~\ref{fig:FIR-fluxes} illustrates how all flux predictions made for $P=8~\mathrm{h}$ scatter and compare with our improved \textit{Herschel}/PACS flux measurements, which are plotted with their respective uncertainties.

\subsection{No flux contribution from a satellite}
\label{sec:noflux}
We consider a thermophysical model as plausible if it predicts far-infrared fluxes that fit to the measured \textit{Herschel}/PACS fluxes with a reduced $\chi^2 \leq 1.7$. Table~\ref{tab:table7} summarizes the parameter space of models that fulfilled this criteria, assuming $P = 8~\mathrm{h}$. 

\begin{table*}[tbp]
  \setlength{\extrarowheight}{2.5pt}
  \centering
  \caption{Results from a parametric study with a TPM code, assuming a rotation period of $P = 8~\mathrm{h}$ and no flux contribution from the satellite.}
  \label{tab:table7}
    \begin{tabular}{cccc}
    \hline\hline
   Parameter  & very low roughness & intermediate roughness & very high roughness \\
                    & (0.1 rms) & (0.5 rms) & (0.9 rms) \\
\hline
    $\theta $ (degrees) &                                                               $ 45 - 90 $ &             $ 60 - 70 $ &           $ 60 - 90 $ \\
    $ \varGamma $ ($\mathrm{J~m^{-2}~s^{-0.5}~K^{-1}}$) &  $ 0.7 - 10 $ &          $ 3 - 10 $ &              $ 3 - 10 $ \\
    $ a = b $ (km) &                                                                    $ 640 - 646$ &            $ 640 - 646 $ &                 $ 640 $ \\
    $ c $ (km) &                                                                                $ 524 - 598$ &            $ 565 - 591 $ &                 $ 565 - 585 $\\
    $ a/c $ &                                                                           $ 1.08 - 1.22$ &  $ 1.09 - 1.13$ &        $ 1.09 - 1.13$ \\
    $ d_\mathrm{Sphere,eff}~(\mathrm{km})$ &                            $ 599 - 629$ &            $ 614 - 627$ &          $ 614 - 621 $ \\
    reduced $\chi^2$ &                                                          $ 1.40 - 1.63$&           $ 1.42 - 1.70$ &        $ 1.46 - 1.66 $\\
    \hline
    \end{tabular}
\end{table*}

We consider the very high roughness case ($s = 0.9~\mathrm{rms}$) as rather unrealistic, as such high roughness values are typically not used in thermophysical simulations. For near-Earth objects, typical roughness values in simulations are $s = 0.2~\mathrm{rms}$, while values above $s > 0.5~\mathrm{rms}$ do not find application. Also, the thermal inertia that is required in this case to produce plausible models is rather high. Recent work by \citet{Lellouch2013} finds evidence that thermal inertia decreases with increasing heliocentric distance, implying a trend towards more and more porous surfaces. According to their study, typical values at heliocentric distances of $r_\mathrm{H} > 41~\mathrm{AU}$ are expected to be around $\varGamma \approx 2~\mathrm{J~m^{-2}~s^{-0.5}~K^{-1}}$, while values do not exceed $\varGamma \approx 6~\mathrm{J~m^{-2}~s^{-0.5}~K^{-1}}$ that far out in the solar system. This was also the reason why we limited our parameter study to thermal inertia values of $\varGamma \leq 10~\mathrm{J~m^{-2}~s^{-0.5}~K^{-1}}$.

For a rotation period of $P = 8~\mathrm{h}$, all plausible solutions cover a range of the principal axis tilt of $\theta = 45 - 90\degree$. The estimated effective diameter (of a sphere with equal volume) is $d_\mathrm{eff} = 599 - 629~\mathrm{km}$. When assuming a rotation period of $P > 8~\mathrm{h}$, it is more difficult to find models that are in agreement with the \textit{Herschel}/PACS measurements. At $P = 20.25~\mathrm{h}$, we obtain $d_\mathrm{eff} = 605 - 625~\mathrm{km}$ at principal axis tilt angles of $\theta = 60 - 90\degree$; solutions can then only be found for low and intermediate surface roughness levels. To explain the size measurement from the occultation under the assumption of a surface with high roughness and a long rotation period, a thermal inertia would be required that is too high to be physically plausible. 

\subsection{Some flux contribution from a satellite}
Given that the broad \textit{HST} F606W band pass is reasonably close to the standard V band pass from the perspective of solar system studies, we translate the satellite's magnitude difference of $\Delta m = 3.79 \pm 0.24~\mathrm{mag}$ into an absolute magnitude of $H_\mathrm{Sat} = 7.48 \pm 0.29~\mathrm{mag}$. This implies that the satellite emits about 3\% of the combined flux. Assuming an identical albedo and using the equation from \citet{Harris1998},
\begin{equation}
d_\mathrm{Sat} = 10^{-\frac{H_\mathrm{V,Sat}}{5}}~1329~p_\mathrm{V}^{-\frac{1}{2}}
,\end{equation} 
we obtain a diameter estimate of $d_\mathrm{Sat} = 112.2 \pm 75.5~\mathrm{km}$ for the satellite. For this scenario, we estimated with our TPM that the satellite would contribute a flux of 0.3~/~0.4~/~0.3~mJy in the \textit{Herschel}/PACS bands (70~/~100~/~160~\textmu m).

Using the same albedo for the satellite and for the primary is an assumption that is often made in literature, but we do not know how realistic it is. For example, the New Horizons mission proved that Charon is significantly darker than Pluto \citep{Buratti2016}. It therefore appears more meaningful to study a worst case scenario for the satellite: We assume that its albedo is lower than for the primary ($p_\mathrm{V,Sat}=5\%$) and that its brightness is at the upper uncertainty limit of the available measurements ($H_\mathrm{V, Sat} = 7.19~\mathrm{mag}$), implying an equivalent diameter of $d_\mathrm{Sat}=217~\mathrm{km}$. Using our TPM, we estimate that this type of satellite produces 1.6~/~1.8~/~1.3~mJy (corresponding to $\approx~14-16\%$ flux) in the \textit{Herschel}/PACS bands if we use typical simulation parameters: An equator-on viewing geometry, a $P = 8~\mathrm{h}$ rotation period (assuming a tidally locked motion), a low thermal inertia of $\varGamma = 1~\mathrm{J~m^{-2}~s^{-0.5}~K^{-1}}$, and an intermediate level of surface roughness of $s = 0.5~\mathrm{rms}$. Acting as a comparison, the NEATM model \citep[]{Harris1998} produces consistent flux estimates of 1.6~/~1.9~/~1.5~mJy for such a satellite using a beaming parameter of $\eta = 1.2$. Subtracting the satellite's worst-case flux estimates from the measured \textit{Herschel}/PACS fluxes provided earlier in Table~\ref{tab:table4} gives us a minimum flux estimate of 9.9~/~11.5~/~6.7~mJy for 2007~UK\textsubscript{126}. 

Thermophysical modeling considering a satellite flux contribution did not lead to meaningful results at any studied rotation period: Although predicted fluxes typically match measurements in the 70~\textmu m and 100~\textmu m bands well, they poorly fit the measurements in the 160~\textmu m band. This is indicated by a much degraded reduced $\chi^2 \geq 3.32$. Table~\ref{tab:table8} lists the range of possible properties of 2007~UK\textsubscript{126} according to models with a goodness of fit in the range of $3.32 \leq \chi^2 \leq 4.0$. Again, a very high roughness and a very high thermal inertia are rather unrealistic. 

\begin{table*}[tbp]
  \setlength{\extrarowheight}{2.5pt}
  \centering
  \caption{Results from a parametric study with a TPM code, considering a maximum plausible flux contribution from the satellite and assuming a rotation period of $P = 8~\mathrm{h}$ for the primary body.}
  \label{tab:table8}
    \begin{tabular}{cccc}
    \hline\hline
   Parameter  & very low roughness      & intermediate roughness        & very high roughness \\
                    & (0.1 rms)                         & (0.5 rms)                     & (0.9 rms) \\
\hline
$ \theta $ (degrees) &                                                          $ 50 - 70 $ &             $ 70 - 90 $ &           $ 80 - 90 $ \\ 
$ \varGamma $ ($\mathrm{J~m^{-2}~s^{-0.5}~K^{-1}}$) &              $ 3 - 10 $ &             $ 5 - 10 $ &            $ 10 $ \\ 
$ a/c $ &                                                                                       $ 1.09 - 1.18 $ &         $ 1.07 - 1.11 $ &       $ 1.09 - 1.10 $ \\
$ d_\mathrm{Sphere,eff} $ (km) &                                                $ 606 - 627 $ &           $ 618 - 637 $ &                 $ 621 $ \\ 
reduced $\chi^2$ &                                                                      $ 3.43 - 3.87 $ &         $ 3.32 - 4.00 $ &       $ 3.47 - 3.56 $ \\
\hline
    \end{tabular}
\end{table*}

Given the very poor fit of our modeled fluxes to \textit{Herschel}/PACS measurements, it is very likely that the flux contribution from the satellite is much less than the derived worst-case contribution, indicating that the discussion in Section~\ref{sec:noflux} is a realistic approximation.

\subsection{Implications}

\begin{figure}[tb]
   \centering
   \includegraphics[scale=0.7]{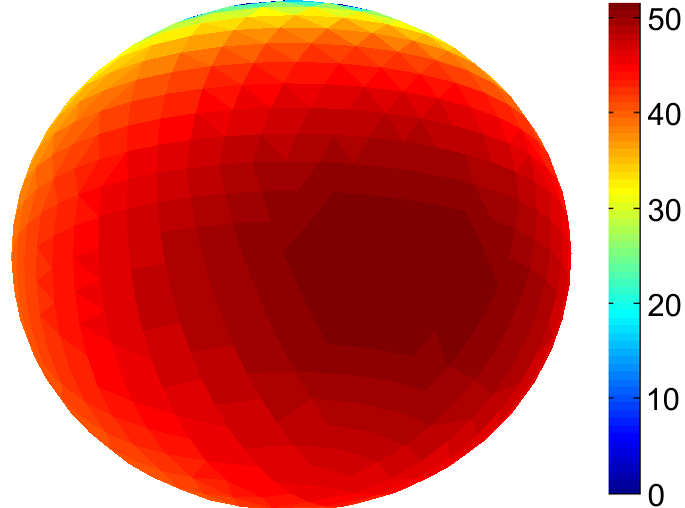}
   \caption{Surface temperature distribution on 2007~UK\textsubscript{126}, as predicted by the TPM for the best model with an intermediate roughness (see Figure~\ref{fig:FIR-fluxes} and text for details). The TPM takes the epoch of the \textit{Herschel} observation into account.}
   \label{fig:temperature-distribution}
    \end{figure}

The TPM simulations enable us to constrain the parameter space marked in Figure~\ref{fig:flattening-ratio} significantly. Figure~\ref{fig:temperature-distribution} illustrates the surface temperature distribution on 2007~UK\textsubscript{126}, as predicted by the TPM for one of the best model fits: A viewing geometry of $\theta=70\degree$ of an oblate sphere with an effective diameter of $d_\mathrm{eff}=618~\mathrm{km}$, a flattening ratio of $a/c = 1.11$, a thermal inertia of $\varGamma = 3~\mathrm{J~m^{-2}~s^{-0.5}~K^{-1}}$, an intermediate level of surface roughness of $s = 0.5~\mathrm{rms}$, and a rotation period of $P = 8~\mathrm{h}$. This fit results in a reduced $\chi^2 = 1.43$ and predicted fluxes of 10.98~/~13.28~/~10.29~mJy in \textit{Herschel's} 70~/~100~/~160~\textmu m bands.

In our range of plausible model fits, we find maximum surface temperatures of about $\approx 50 - 55~\mathrm{K}$ in the subsolar point. Since 2007~UK\textsubscript{126} is still on its way to perihelion in 2046, it spends a large fraction of its orbit at comparable or even higher surface temperatures. From two perspectives, it is unlikely that 2007~UK\textsubscript{126} has retained volatile ices (CH\textsubscript{4}, N\textsubscript{2}, and CO): Because of its estimated temperature levels and  its small size. At the given surface temperatures, all volatiles would be lost solely based on Jeans escape, as indicated by a greatly simplified model by \citet{Schaller2007}. However, as discussed by \citet{Stern2008}, an atmosphere on a body as small as 2007~UK\textsubscript{126} would be governed by hydrodynamic escape (owing to the body's low gravity, even considering moderate to high densities), or by a combination of Jeans and hydrodynamic escape. It is unlikely that Jeans escape alone, the slowest loss mechanism, is depleting the atmosphere. Given the eccentricity of the orbit of 2007~UK\textsubscript{126}, the proportion of each escape mechanism could vary over time, and seasonal freeze out and sublimation of volatile ices could play a role. In addition to classic escape mechanisms, impacts that almost certainly occurred throughout the lifetime of the solar system would accelerate the escape of an atmosphere. The lack of volatiles is supported by available near-infrared spectra (see Section~\ref{sec:properties}) that could not identify any ice features (within the S/N limits of the dataset). However, small amounts of involatile amorphous water ice, below the detection limit of the available spectra, cannot be excluded since they could have been preserved at the calculated surface temperatures.
Since no volatiles, and hence no atmosphere are expected on 2007~UK\textsubscript{126}, the approach of fitting square-well profiles to the occultation light curve data as discussed in Section~\ref{sec:squarewellfits} is reconfirmed.


\section{Conclusions}
For the first time, it was possible to measure the size of a detached object of the TNO population directly during a stellar occultation at three locations. Our findings from the occultation measurements are:

\begin{enumerate}
\item{The shadow ellipse that fits best to the times of disappearance and reappearance on each chord has a major axis of $a_\mathrm{El} = 645.80 \pm 5.68 ~\mathrm{km}$ and a minor axis of $b_\mathrm{El} = 597.81 \pm 12.74~\mathrm{km}$ ($a_\mathrm{El}/b_\mathrm{El} = 1.080 \pm 0.025$)}.
\item{The estimated projected area, combined with previous photometric measurements, suggests an albedo estimate of $p_\mathrm{V} = 15.0 \pm 1.6\%$. This estimate is consistent with and more accurate than an earlier estimate that was based on a size derived solely from a thermophysical model.}
\item{Owing to the very low light curve amplitude and rotation period reported in literature and the size estimated from the occultation, we assume by analogy to other solar system bodies that 2007~UK\textsubscript{126} resembles an oblate spheroid. }
\item{Given the shadow ellipse geometry and the non-zero light curve amplitude, 2007~UK\textsubscript{126} was not seen pole-on.}
\item{We cannot constrain a bulk density at this time. Purely by geometric means, we can constrain the effective diameter of 2007~UK\textsubscript{126} to $d_\mathrm{eff} = 535 - 638~\mathrm{km}$. We argue qualitatively that the spheroid's flattening ratio is likely $a/c \leq 1.4870$, which would constrain the effective diameter range slightly more to $d_\mathrm{eff} = 563 - 638~\mathrm{km}$.}
\item{No secondary occultation has been detected, so the satellite could not be reobserved (although chances would have been extremely small). This implies that it was either behind or in front of the primary, not in the primary's shadow path, or so small that its shadow moved in between the observers' locations. With sufficient distance to the primary at the time of the occultation, the shadow of the secondary may not have even hit the Earth's surface.}
\end{enumerate}

Modeling the thermal emission of 2007~UK\textsubscript{126}, using geometrical constraints obtained from the occultation, and comparing modeled fluxes with improved \textit{Herschel}/PACS far-infrared flux measurements, we conclude that:

\begin{enumerate}
\item{The effective diameter of 2007~UK\textsubscript{126} is likely of the order of $d_\mathrm{eff} = 599 - 629~\mathrm{km}$. The flattening ratio is of the order of $a/c = 1.08 - 1.22$}.
\item{The body was likely seen near an equator-on viewing geometry ($\theta = 45-90\degree$). }
\item{Assuming an albedo of 5\% for the satellite and a worst-case photometric error, its effective diameter would be $217~\mathrm{km}$, from which we derive the worst-case satellite flux contribution in the \textit{Herschel}/PACS bands. Considering a flux contribution from the satellite, it was not possible to derive models at any given rotation period that would be compatible with the occultation shadow size and \textit{Herschel}/PACS measurements, so it is very likely that any satellite flux contributions would be much less.} 
\item{Our models result in maximum surface temperatures of \mbox{$\approx 50 - 55~\mathrm{K}$} in the subsolar point.}
\end{enumerate}

Combining our analyses of the occultation and the thermal emission, we conclude:

\begin{enumerate}
\item{While a true flattening ratio close to bifurcation cannot formally be ruled out from the occultation data, given the unknown density, high flattening ratios can be ruled out thanks to our thermal simulations. This constrains the size of the body considerably.}
\item{Owing to its estimated surface temperature levels and its small size, it is unlikely that 2007~UK\textsubscript{126} has retained volatile ices (CH\textsubscript{4}, N\textsubscript{2}, and CO), which is also supported by its featureless near-infrared spectra reported in literature. This reconfirms fitting square-well profiles to the occultation light curve, as no atmosphere is expected.}
\end{enumerate}

The estimated size supports the status of 2007~UK\textsubscript{126} as a dwarf planet candidate. While faster sampling of an occultation would help to decrease the uncertainties of the shadow ellipse fit, it would not help to significantly constrain the tilt angle $\theta$ of the rotation axis, which is the dominating source of uncertainty for a geometrically derived size estimate. Without a density estimate, only the observation of multiple occultations, extensive light curve measurements, and work on a shape model could eventually provide better geometrical constraints on the body's true size. Thermophysical modeling allows to overcome this issue. Future occultations of 2007~UK\textsubscript{126} or acquisition of additional high-resolution image data will hopefully provide opportunities to reobserve its satellite. Knowledge of the satellite's size and orbit is crucial to constrain the system mass and bulk density, which would immediately constrain plausible shapes. While size and albedo estimates solely derived from far-infrared measurements still have considerable uncertainties, the combined analysis of occultation observations and far-infrared measurements is a powerful tool to thoroughly characterize bodies in our solar system. 

\begin{acknowledgements}
For their efforts on long-term occultation predictions and updates, we thank B.~Sicardy (LESIA, Observatoire de Paris), J.~Desmars (IMCEE, Paris), N.~Morales, P.~Santos-Sanz, R.~Duffard (Instituto de Astrofisica de Andalucia-CSIC) and the following members from the Rio Group: R.~Vieira-Martins, M.~Assafin, A.~Dias-Oliveira, B.~Morgado, G.~Benedetti-Rossi and A.~Ramos Gomes Jr.. We would like to thank M.~H\"ummer for his work on creating a video sequence via ray tracing illustrating the occultation based on the reconstructed geometry of 2007~UK\textsubscript{126} (see supplementary material to this article). K.S.~has been supported financially by a one-year fellowship from the German Academic Exchange Service (DAAD) during his stay at NASA Ames Research Center, CA as part of his PhD research. K.S.~would also like to thank the Universities Space Research Association (USRA) for their support throughout his thesis. T.M., C.K.~and J.L.O.~have received funding from the European Union's Horizon 2020 Research and Innovation Programme, under Grant Agreement No. 687378. C.K.~has been supported by the PECS grant \# 4000109997/13/NL/KML of the Hungarian Space Office and the European Space Agency, the K-104607 grant of the Hungarian Research Fund (OTKA), and the GINOP-2.3.2-15-2016-00003 grant of the National Research, Development and Innovation Office of Hungary (NKFIH). J.L.O.~acknowledges support from Proyecto de Excelencia J.A. 2012-FQM1776. The Astronomical Telescope of the University Stuttgart (ATUS) has been funded in parts by the German Aerospace Center (DLR). K.S.~and J.W.~would like to thank all involved parties at Sierra Remote Observatories (SRO) for providing and maintaining the infrastructure that allows us to remotely operate ATUS. L.~Van~Vleet was manually monitoring and controlling the observatory roof that night which allowed us to conduct this observation despite of critical humidity levels. K.S.~thanks T.~George and R.L.~Anderson, the developers of R-OTE, for vivid discussions on statistical methods and camera- and VTI-specific time offset corrections. This research has made use of NASA's Astrophysics Data System.
\end{acknowledgements}


\bibliographystyle{aa}
\bibliography{References}

\begin{thebibliography}{76}
\expandafter\ifx\csname natexlab\endcsname\relax\def\natexlab#1{#1}\fi

\bibitem[{{Balog} {et~al.}(2014){Balog}, {M\"uller}, {Nielbock}, {Altieri},
  {Klaas}, {Blommaert}, {Linz}, {Lutz}, {Mo\'or}, {Billot}, {Sauvage}, \&
  {Okumura}}]{Balog2014}
{Balog}, Z., {M\"uller}, T., {Nielbock}, M., {et~al.} 2014, Exp. Astron., 37,
  129

\bibitem[{{Barucci} {et~al.}(2011){Barucci}, {Alvarez-Candal}, {Merlin},
  {Belskaya}, {de Bergh}, {Perna}, {DeMeo}, \& {Fornasier}}]{Barucci2011}
{Barucci}, M.~A., {Alvarez-Candal}, A., {Merlin}, F., {et~al.} 2011, \icarus,
  214, 297

\bibitem[{{Barucci} {et~al.}(2005){Barucci}, {Belskaya}, {Fulchignoni}, \&
  {Birlan}}]{Barucci2005}
{Barucci}, M.~A., {Belskaya}, I.~N., {Fulchignoni}, M., \& {Birlan}, M. 2005,
  \aj, 130, 1291

\bibitem[{{Bohlin} \& {Gilliland}(2004)}]{Bohlin2004}
{Bohlin}, R.~C. \& {Gilliland}, R.~L. 2004, \aj, 127, 3508

\bibitem[{{Bosh} {et~al.}(2016){Bosh}, {Person}, {Zuluaga}, {more}, {more}, \&
  {even more}}]{Bosh2016}
{Bosh}, A.~S., {Person}, M.~J., {Zuluaga}, C.~A., {et~al.} 2016, {Icarus
  (submitted)}

\bibitem[{{Braga-Ribas} {et~al.}(2013){Braga-Ribas}, {Sicardy}, {Ortiz},
  {Lellouch}, {Tancredi}, {Lecacheux}, {Vieira-Martins}, {Camargo}, {Assafin},
  {Behrend}, {Vachier}, {Colas}, {Morales}, {Maury}, {Emilio}, {Amorim},
  {Unda-Sanzana}, {Roland}, {Bruzzone}, {Almeida}, {Rodrigues}, {Jacques},
  {Gil-Hutton}, \& more}]{Braga-Ribas2013}
{Braga-Ribas}, F., {Sicardy}, B., {Ortiz}, J.~L., {et~al.} 2013, \apj, 773, 26

\bibitem[{{Braga-Ribas} {et~al.}(2014{\natexlab{a}}){Braga-Ribas}, {Sicardy},
  {Ortiz}, {Snodgrass}, {Roques}, {Vieira-Martins}, {Camargo}, {Assafin},
  {Duffard}, {Jehin}, {Pollock}, {Leiva}, {Emilio}, {Machado}, {Colazo},
  {Lellouch}, {Skottfelt}, {Gillon}, {Ligier}, {Maquet}, {Benedetti-Rossi},
  {Gomes}, {Kervella}, {Monteiro}, {Sfair}, {El Moutamid}, {Tancredi},
  {Spagnotto}, {Maury}, {Morales}, {Gil-Hutton}, {Roland}, {Ceretta}, {Gu},
  {Wang}, {Harps{\o}e}, {Rabus}, {Manfroid}, {Opitom}, {Vanzi}, {Mehret},
  {Lorenzini}, {Schneiter}, {Melia}, {Lecacheux}, {Colas}, {Vachier},
  {Widemann}, {Almenares}, {Sandness}, {Char}, {Perez}, {Lemos}, {Martinez},
  {J{\o}rgensen}, {Dominik}, {Roig}, {Reichart}, {Lacluyze}, {Haislip},
  {Ivarsen}, {Moore}, {Frank}, \& {Lambas}}]{Braga-Ribas2014a}
{Braga-Ribas}, F., {Sicardy}, B., {Ortiz}, J.~L., {et~al.} 2014{\natexlab{a}},
  \nat, 508, 72

\bibitem[{{Braga-Ribas} {et~al.}(2014{\natexlab{b}}){Braga-Ribas},
  {Vieira-Martins}, {Assafin}, {Camargo}, {Sicardy}, \&
  {Ortiz}}]{Braga-Ribas2014b}
{Braga-Ribas}, F., {Vieira-Martins}, R., {Assafin}, M., {et~al.}
  2014{\natexlab{b}}, in {Revista Mexicana de Astronomia y Astrofisica
  Conference Series}, Vol.~44, 3--3

\bibitem[{{Brown}(2013{\natexlab{a}})}]{Brown2013}
{Brown}, M.~E. 2013{\natexlab{a}}, ApJL, 767, L7

\bibitem[{{Brown}(2013{\natexlab{b}})}]{Brown2013b}
{Brown}, M.~E. 2013{\natexlab{b}}, ApJL, 778, L34

\bibitem[{{Buratti} {et~al.}(2016){Buratti}, {Hofgartner}, {Hicks}, {Weaver},
  {Stern}, {Momary}, {Mosher}, {Beyer}, {Young}, {Ennico}, \&
  {Olkin}}]{Buratti2016}
{Buratti}, B.~J., {Hofgartner}, J.~D., {Hicks}, M.~D., {et~al.} 2016, {ArXiv
  e-prints} [\eprint[arXiv]{1604.06129}]

\bibitem[{{Camargo} {et~al.}(2014){Camargo}, {Vieira-Martins}, {Assafin},
  {Braga-Ribas}, {Sicardy}, {Desmars}, {Andrei}, {Benedetti-Rossi}, \&
  {Dias-Oliveira}}]{Camargo2014}
{Camargo}, J. I.~B., {Vieira-Martins}, R., {Assafin}, M., {et~al.} 2014, \aap,
  561, A37

\bibitem[{{Chandrasekhar}(1967)}]{Chandrasekhar1967}
{Chandrasekhar}, S. 1967, Comm. Pure Appl. Math., 20, 251

\bibitem[{{Collins} \& {Kielkopf}(2013)}]{Collins2013}
{Collins}, K. \& {Kielkopf}, J. 2013 [\eprint[ascl]{1309.001}]

\bibitem[{{Collins}(2015)}]{Collins2015}
{Collins}, K.~A. 2015, PhD thesis, {University of Louisville}

\bibitem[{{Collins} {et~al.}(2016){Collins}, {Kielkopf}, \&
  {Stassun}}]{Collins2016}
{Collins}, K.~A., {Kielkopf}, J.~F., \& {Stassun}, K.~G. 2016, {ArXiv e-prints}
  [\eprint[arXiv]{1601.02622}]

\bibitem[{{Descamps} {et~al.}(2011){Descamps}, {Marchis}, {Berthier}, {Emery},
  {Duch{\^e}ne}, {de Pater}, {Wong}, {Lim}, {Hammel}, {Vachier}, {Wiggins},
  {Teng-Chuen-Yu}, {Peyrot}, {Pollock}, {Assafin}, {Vieira-Martins}, {Camargo},
  {Braga-Ribas}, \& {Macomber}}]{Descamps2011}
{Descamps}, P., {Marchis}, F., {Berthier}, J., {et~al.} 2011, \icarus, 211,
  1022

\bibitem[{{Duffard} {et~al.}(2009){Duffard}, {Ortiz}, {Thirouin},
  {Santos-Sanz}, \& {Morales}}]{Duffard2009}
{Duffard}, R., {Ortiz}, J.~L., {Thirouin}, A., {Santos-Sanz}, P., \& {Morales},
  N. 2009, \aap, 505, 1283

\bibitem[{{Elliot} {et~al.}(2010){Elliot}, {Person}, {Zuluaga}, {Bosh},
  {Adams}, {Brothers}, {Gulbis}, {Levine}, {Lockhart}, {Zangari}, {Babcock},
  {Dupr{\'e}}, {Pasachoff}, {Souza}, {Rosing}, {Secrest}, {Bright}, {Dunham},
  {Sheppard}, {Kakkala}, {Tilleman}, {Berger}, {Briggs}, {Jacobson}, {Valleli},
  {Volz}, {Rapoport}, {Hart}, {Brucker}, {Michel}, {Mattingly},
  {Zambrano-Marin}, {Meyer}, {Wolf}, {Ryan}, {Ryan}, {Morzinski}, {Grigsby},
  {Brimacombe}, {Ragozzine}, {Montano}, \& {Gilmore}}]{Elliot2010}
{Elliot}, J.~L., {Person}, M.~J., {Zuluaga}, C.~A., {et~al.} 2010, \nat, 465,
  897

\bibitem[{{Engineering Industries Association (EIA)}(1957)}]{EIA1957}
{Engineering Industries Association (EIA)}. 1957, {Standard RS 170, Electrical
  Performance Standards - Monochrome Television Studio Facilities}

\bibitem[{{Fornasier} {et~al.}(2009){Fornasier}, {Barucci}, {de Bergh},
  {Alvarez-Candal}, {DeMeo}, {Merlin}, {Perna}, {Guilbert}, {Delsanti},
  {Dotto}, \& {Doressoundiram}}]{Fornasier2009}
{Fornasier}, S., {Barucci}, M.~A., {de Bergh}, C., {et~al.} 2009, \aap, 508,
  457

\bibitem[{{Fornasier} {et~al.}(2013){Fornasier}, {Lellouch}, {M\"uller},
  {Santos-Sanz}, {Panuzzo}, {Kiss}, {Lim}, {Mommert}, {Bockel\'ee-Morvan},
  {Vilenius}, {Stansberry}, {Tozzi}, {Mottola}, {Delsanti}, {Crovisier},
  {Duffard}, {Henry}, {Lacerda}, {Barucci}, \& {Gicquel}}]{Fornasier2013}
{Fornasier}, S., {Lellouch}, E., {M\"uller}, T., {et~al.} 2013, \aap, 555, A15

\bibitem[{{Fukugita} {et~al.}(1995){Fukugita}, {Shimasaku}, \&
  {Ichikawa}}]{Fukugita1995}
{Fukugita}, M., {Shimasaku}, K., \& {Ichikawa}, T. 1995, \pasp, 107, 945

\bibitem[{{Fulchignoni} {et~al.}(2008){Fulchignoni}, {Belskaya}, {Barucci}, {de
  Sanctis}, \& {Doressoundiram}}]{Fulchignoni2008}
{Fulchignoni}, M., {Belskaya}, I., {Barucci}, M.~A., {de Sanctis}, M.~C., \&
  {Doressoundiram}, A. 2008, in {The Solar System Beyond Neptune}, ed. M.~A.
  {Barucci}, H.~{Boehnhardt}, D.~P. {Cruikshank}, A.~{Morbidelli}, \&
  R.~{Dotson} ({The University of Arizona Press}), 181--192

\bibitem[{{Gaia Collaboration} {et~al.}(2016){Gaia Collaboration}, {Brown},
  {Vallenari}, {Prusti}, {de Bruijne}, {Mignard}, {Drimmel}, \&
  {co-authors}}]{GAIA2016}
{Gaia Collaboration}, {Brown}, A.~G.~A., {Vallenari}, A., {et~al.} 2016, {ArXiv
  e-prints} [\eprint[arXiv]{1609.04172}]

\bibitem[{{George}(2014)}]{George2014}
{George}, T. 2014, {Analysis of Camera Delay Corrections for use in R-OTE and
  Occular Occultation Timing Extraction Programs}, {White Paper}

\bibitem[{{George} \& {Anderson}(2013)}]{George2013}
{George}, T. \& {Anderson}, R.~L. 2013, in {2013 IOTA Annual Meeting, Toronto,
  Canada}

\bibitem[{{Gladman} {et~al.}(2008){Gladman}, {Marsden}, \&
  {Vanlaerhoven}}]{Gladman2008}
{Gladman}, B., {Marsden}, B.~G., \& {Vanlaerhoven}, C. 2008, in {The Solar
  System Beyond Neptune}, ed. M.~A. {Barucci}, H.~{Boehnhardt}, D.~P.
  {Cruikshank}, A.~{Morbidelli}, \& R.~{Dotson} ({The University of Arizona
  Press}), 43--57

\bibitem[{{Harris}(1994)}]{Harris1994}
{Harris}, A.~W. 1994, \icarus, 107, 209

\bibitem[{{Harris}(1998)}]{Harris1998}
{Harris}, A.~W. 1998, \icarus, 131, 291

\bibitem[{{Herald}(2016)}]{OccultHelp}
{Herald}, D. 2016, {Occult v4 Help}

\bibitem[{{Jacobson} {et~al.}(1992){Jacobson}, {Campbell}, {Taylor}, \&
  {Synnott}}]{Jacobson1992}
{Jacobson}, R.~A., {Campbell}, J.~K., {Taylor}, A.~H., \& {Synnott}, S.~P.
  1992, \aj, 103, 2068

\bibitem[{{Johnston}(2014)}]{Johnston2014}
{Johnston}, W.~R. 2014, {TNO and Centaur Diameters, Albedo, and Densities V2.0,
  EAR-A-COMPIL-5-TNOCENALB-V2.0, NASA Planetary Data System}

\bibitem[{{Kiss} {et~al.}(2014){Kiss}, {M\"uller}, {Vilenius}, {P\'al},
  {Santos-Sanz}, {Lellouch}, {Marton}, {Vereb\'elyi}, {Szalai}, {Hartogh},
  {Stansberry}, {Henry}, \& {Delsanti}}]{Kiss2014}
{Kiss}, C., {M\"uller}, T., {Vilenius}, E., {et~al.} 2014, Exp. Astron., 37,
  161

\bibitem[{{Lacerda} {et~al.}(2014){Lacerda}, {Fornasier}, {Lellouch}, {Kiss},
  {Vilenius}, {Santos-Sanz}, {Rengel}, {M\"uller}, {Stansberry}, {Duffard},
  {Delsanti}, \& {Guilbert-Lepoutre}}]{Lacerda2014}
{Lacerda}, P., {Fornasier}, S., {Lellouch}, E., {et~al.} 2014, ApJL, 793, L2

\bibitem[{{Lellouch} {et~al.}(2013){Lellouch}, {Santos-Sanz}, {Lacerda},
  {Mommert}, {Duffard}, {Ortiz}, {M\"uller}, {Fornasier}, {Stansberry}, {Kiss},
  {Vilenius}, {Mueller}, {Peixinho}, {Moreno}, {Groussin}, {Delsanti}, \&
  {Harris}}]{Lellouch2013}
{Lellouch}, E., {Santos-Sanz}, P., {Lacerda}, P., {et~al.} 2013, \aap, 557, A60

\bibitem[{{Lindegren} {et~al.}(2016){Lindegren}, {Lammers}, {Bastian},
  {Hern{\'a}ndez}, {Klioner}, {Hobbs}, {Bombrun}, {Michalik}, {Ramos-Lerate},
  {Butkevich}, {Comoretto}, {Joliet}, {Holl}, {Hutton}, {Parsons},
  {Steidelm{\"u}ller}, {Abbas}, {Altmann}, {Andrei}, {Anton}, {Bach},
  {Barache}, {Becciani}, {Berthier}, {Bianchi}, {Biermann}, {Bouquillon},
  {Bourda}, {Br{\"u}semeister}, {Bucciarelli}, {Busonero}, {Carlucci},
  {Casta{\~n}eda}, {Charlot}, {Clotet}, {Crosta}, {Davidson}, {de Felice},
  {Drimmel}, {Fabricius}, {Fienga}, {Figueras}, {Fraile}, {Gai}, {Garralda},
  {Geyer}, {Gonz{\'a}lez-Vidal}, {Guerra}, {Hambly}, {Hauser}, {Jordan},
  {Lattanzi}, {Lenhardt}, {Liao}, {L{\"o}ffler}, {McMillan}, {Mignard}, {Mora},
  {Morbidelli}, {Portell}, {Riva}, {Sarasso}, {Serraller}, {Siddiqui}, {Smart},
  {Spagna}, {Stampa}, {Steele}, {Taris}, {Torra}, {van Reeven}, {Vecchiato},
  {Zschocke}, {de Bruijne}, {Gracia}, {Raison}, {Lister}, {Marchant},
  {Messineo}, {Soffel}, {Osorio}, {de Torres}, \& {O'Mullane}}]{Lindegren2016}
{Lindegren}, L., {Lammers}, U., {Bastian}, U., {et~al.} 2016, ArXiv e-prints
  [\eprint[arXiv]{1609.04303}]

\bibitem[{{Millis} \& {Dunham}(1989)}]{Millis1989}
{Millis}, R.~L. \& {Dunham}, D.~W. 1989, in {Asteroid II}, ed. R.~P. {Binzel},
  T.~{Gehrels}, \& M.~S. {Matthews} ({University of Arizona Press}), 148--170

\bibitem[{{Millis} \& {Elliot}(1979)}]{Millis1979}
{Millis}, R.~L. \& {Elliot}, J.~L. 1979, in {Asteroids}, ed. T.~{Gehrels}
  ({University of Arizona Press}), 98--118

\bibitem[{{Mommert} {et~al.}(2012){Mommert}, {Harris}, {Kiss}, {P\'al},
  {Santos-Sanz}, {Stansberry}, {Delsanti}, {Vilenius}, {Mueller}, {Peixinho},
  {Lellouch}, {Szalai}, {Henry}, {Duffard}, {Fornasier}, {Hartogh}, {Mueller},
  {Ortiz}, {Protopapa}, {Rengel}, \& {Thirouin}}]{Mommert2012}
{Mommert}, M., {Harris}, A.~W., {Kiss}, C., {et~al.} 2012, \aap, 541, A93

\bibitem[{{M\"uller} \& {Lagerros}(1998)}]{Mueller1998}
{M\"uller}, T.~G. \& {Lagerros}, J. S.~V. 1998, \aap, 338, 340

\bibitem[{{M\"uller} \& {Lagerros}(2002)}]{Mueller2002}
{M\"uller}, T.~G. \& {Lagerros}, J. S.~V. 2002, \aap, 381, 324

\bibitem[{{M\"uller} {et~al.}(2009){M\"uller}, {Lellouch}, {B\"ohnhardt},
  {Stansberry}, {Barucci}, {Crovisier}, {Delsanti}, {Doressoundiram}, {Dotto},
  {Duffard}, {Fornasier}, {Groussin}, {Guti\'errez}, {Hainaut}, {Harris},
  {Hartogh}, {Hestroffer}, {Horner}, {Jewitt}, {Kidger}, {Kiss}, {Lacerda},
  {Lara}, {Lim}, {Mueller}, {Moreno}, {Ortiz}, {Rengel}, {Santos-Sanz},
  {Swinyard}, {Thomas}, {Thirouin}, \& {Trilling}}]{Mueller2009}
{M\"uller}, T.~G., {Lellouch}, E., {B\"ohnhardt}, H., {et~al.} 2009, EM\&P,
  105, 209

\bibitem[{{Noll} {et~al.}(2009){Noll}, {Grundy}, {Benecchi}, {Levison}, \&
  {Barker}}]{Noll2009}
{Noll}, K.~S., {Grundy}, W.~M., {Benecchi}, S.~D., {Levison}, H.~F., \&
  {Barker}, E.~A. 2009, in {AAS/Division for Planetary Sciences Meeting
  Abstracts}, Vol.~41, 47.07

\bibitem[{{Ortiz} {et~al.}(2012){Ortiz}, {Sicardy}, {Braga-Ribas},
  {Alvarez-Candal}, {Lellouch}, {Duffard}, {Pinilla-Alonso}, {Ivanov},
  {Littlefair}, {Camargo}, {Assafin}, {Unda-Sanzana}, {Jehin}, {Morales},
  {Tancredi}, {Gil-Hutton}, {de La Cueva}, {Colque}, {da Silva Neto},
  {Manfroid}, {Thirouin}, {Guti{\'e}rrez}, {Lecacheux}, {Gillon}, {Maury},
  {Colas}, {Licandro}, {Mueller}, {Jacques}, {Weaver}, {Milone}, {Salvo},
  {Bruzzone}, {Organero}, {Behrend}, {Roland}, {Vieira-Martins}, {Widemann},
  {Roques}, {Santos-Sanz}, {Hestroffer}, {Dhillon}, {Marsh}, {Harlingten},
  {Bagatin}, {Alonso}, {Ortiz}, {Colazo}, {Lima}, {Oliveira}, {Kerber},
  {Smiljanic}, {Pimentel}, {Giacchini}, {Cacella}, \& {Emilio}}]{Ortiz2012}
{Ortiz}, J.~L., {Sicardy}, B., {Braga-Ribas}, F., {et~al.} 2012, \nat, 491, 566

\bibitem[{{Ortiz} {et~al.}(2014){Ortiz}, {Sicardy}, {Braga-Ribas}, {Morales},
  {Duffard}, {Santos-Sanz}, {Vieira-Martins}, {Camargo}, {Assafin}, {Roques},
  {Widemann}, {Lecacheux}, \& {Colas}}]{Ortiz2014}
{Ortiz}, J.~L., {Sicardy}, B., {Braga-Ribas}, F., {et~al.} 2014, in {European
  Planetary Science Congress 2014 Abstracts}, Vol.~9, EPSC2014--525

\bibitem[{{P{\"a}tzold} {et~al.}(2016){P{\"a}tzold}, {Andert}, {Hahn}, {Asmar},
  {Barriot}, {Bird}, {H{\"a}usler}, {Peter}, {Tellmann}, {Gr{\"u}n},
  {Weissman}, {Sierks}, {Jorda}, {Gaskell}, {Preusker}, \&
  {Scholten}}]{Paetzold2016}
{P{\"a}tzold}, M., {Andert}, T., {Hahn}, M., {et~al.} 2016, \nat, 530, 63

\bibitem[{{Pavlov}(2014)}]{Pavlov2014}
{Pavlov}, H. 2014, in {Eighth Trans-Tasman Symposium on Occultations (TTSO8),
  Melbourne, Australia}

\bibitem[{{Perna} {et~al.}(2010){Perna}, {Barucci, M. A.}, {Fornasier, S.},
  {DeMeo, F. E.}, {Alvarez-Candal, A.}, {Merlin, F.}, {Dotto, E.},
  {Doressoundiram, A.}, \& {de Bergh, C.}}]{Perna2010}
{Perna}, D., {Barucci, M. A.}, {Fornasier, S.}, {et~al.} 2010, \aap, 510, A53

\bibitem[{{Person} {et~al.}(2013){Person}, {Dunham}, {Bosh}, {Levine},
  {Gulbis}, {Zangari}, {Zuluaga}, {Pasachoff}, Babcock, Pandey, Amrhein,
  Sallum, Tholen, Collins, Bida, Taylor, Bright, Wolf, Meyer, Pfueller,
  Wiedemann, Roeser, Lucas, Kakkala, Ciotti, Plunkett, Hiraoka, Best, Pilger,
  Micheli, Springmann, Hicks, Thackeray, Emery, Tilleman, Harris, Sheppard,
  Rapoport, Ritchie, Pearson, Mattingly, Brimacombe, Gault, Jones, Nolthenius,
  Broughton, \& Barry}]{Person2013}
{Person}, M.~J., {Dunham}, E.~W., {Bosh}, A.~S., {et~al.} 2013, \aj, 146, 83

\bibitem[{{Pf\"uller} {et~al.}(2012){Pf\"uller}, {Wolf}, {Hall}, \&
  {R\"oser}}]{Pfueller2012}
{Pf\"uller}, E., {Wolf}, J., {Hall}, H., \& {R\"oser}, H.-P. 2012, \procspie,
  8444, 844413

\bibitem[{{Pilbratt} {et~al.}(2010){Pilbratt}, {Riedinger}, {Passvogel},
  {Crone}, {Doyle}, {Gageur}, {Heras}, {Jewell}, {Metcalfe}, {Ott}, \&
  {Schmidt}}]{Pilbratt2010}
{Pilbratt}, G.~L., {Riedinger}, J.~R., {Passvogel}, T., {et~al.} 2010, \aap,
  518, L1

\bibitem[{{Poglitsch} {et~al.}(2010){Poglitsch}, {Waelkens}, {Geis},
  {Feuchtgruber}, {Vandenbussche}, {Rodriguez, L.}, {Krause, O.}, {Renotte,
  E.}, {van Hoof, C.}, {Saraceno, P.}, {Cepa, J.}, {Kerschbaum, F.}, {Agn\`ese,
  P.}, {Ali, B.}, {Altieri, B.}, {Andreani, P.}, {Augueres, J.-L.}, {Balog,
  Z.}, {Barl, L.}, {Bauer, O. H.}, {Belbachir, N.}, {Benedettini, M.}, {Billot,
  N.}, {Boulade, O.}, {Bischof, H.}, {Blommaert, J.}, {Callut, E.}, {Cara, C.},
  {Cerulli, R.}, {Cesarsky, D.}, {Contursi, A.}, {Creten, Y.}, {De Meester,
  W.}, {Doublier, V.}, {Doumayrou, E.}, {Duband, L.}, {Exter, K.}, {Genzel,
  R.}, {Gillis, J.-M.}, {Gr\"ozinger, U.}, {Henning, T.}, {Herreros, J.},
  {Huygen, R.}, {Inguscio, M.}, {Jakob, G.}, {Jamar, C.}, {Jean, C.}, {de Jong,
  J.}, {Katterloher, R.}, {Kiss, C.}, {Klaas, U.}, {Lemke, D.}, {Lutz, D.},
  {Madden, S.}, {Marquet, B.}, {Martignac, J.}, {Mazy, A.}, {Merken, P.},
  {Montfort, F.}, {Morbidelli, L.}, {M\"uller, T.}, {Nielbock, M.}, {Okumura,
  K.}, {Orfei, R.}, {Ottensamer, R.}, {Pezzuto, S.}, {Popesso, P.}, {Putzeys,
  J.}, {Regibo, S.}, {Reveret, V.}, {Royer, P.}, {Sauvage, M.}, {Schreiber,
  J.}, {Stegmaier, J.}, {Schmitt, D.}, {Schubert, J.}, {Sturm, E.}, {Thiel,
  M.}, {Tofani, G.}, {Vavrek, R.}, {Wetzstein, M.}, {Wieprecht, E.}, \&
  {Wiezorrek, E.}}]{Poglitsch2010}
{Poglitsch}, A., {Waelkens}, C., {Geis}, N., {et~al.} 2010, \aap, 518, L2

\bibitem[{{Roatsch} {et~al.}(2009){Roatsch}, {Jaumann}, {Stephan}, \&
  {Thomas}}]{Roatsch2009}
{Roatsch}, T., {Jaumann}, R., {Stephan}, K., \& {Thomas}, P. 2009, in {Saturn
  from Cassini-Huygens}, ed. M.~Dougherty, L.~Esposito, \& S.~Krimigis
  ({Springer Netherlands}), 763--781

\bibitem[{{Roques} {et~al.}(1987){Roques}, {Moncuquet}, \&
  {Sicardy}}]{Roques1987}
{Roques}, F., {Moncuquet}, M., \& {Sicardy}, B. 1987, \aj, 93, 1549

\bibitem[{{Santos-Sanz} {et~al.}(2012){Santos-Sanz}, {Lellouch}, {Fornasier},
  {Kiss}, {P\'al}, {M\"uller, T. G.}, {Vilenius, E.}, {Stansberry, J.},
  {Mommert, M.}, {Delsanti, A.}, {Mueller, M.}, {Peixinho, N.}, {Henry, F.},
  {Ortiz, J. L.}, {Thirouin, A.}, {Protopapa, S.}, {Duffard, R.}, {Szalai, N.},
  {Lim, T.}, {Ejeta, C.}, {Hartogh, P.}, {Harris, A. W.}, \& {Rengel,
  M.}}]{Santos-Sanz2012}
{Santos-Sanz}, P., {Lellouch}, E., {Fornasier}, S., {et~al.} 2012, \aap, 541,
  A92

\bibitem[{{Schaller} \& {Brown}(2007)}]{Schaller2007}
{Schaller}, E.~L. \& {Brown}, M.~E. 2007, ApJL, 659, L61

\bibitem[{{Schneider} {et~al.}(2012){Schneider}, {Rasband}, \&
  {Eliceiri}}]{Schneider2012}
{Schneider}, C.~A., {Rasband}, W.~S., \& {Eliceiri}, K.~W. 2012, Nat. Methods,
  671

\bibitem[{{Sheppard} \& {Jewitt}(2002)}]{Sheppard2002}
{Sheppard}, S.~S. \& {Jewitt}, D.~C. 2002, \aj, 124, 1757

\bibitem[{{Sicardy} {et~al.}(2011){Sicardy}, {Ortiz}, {Assafin}, {Jehin},
  {Maury}, {Lellouch}, {Hutton}, {Braga-Ribas}, {Colas}, {Hestroffer},
  {Lecacheux}, {Roques}, {Santos-Sanz}, {Widemann}, {Morales}, {Duffard},
  {Thirouin}, {Castro-Tirado}, {Jel{\'{\i}}nek}, {Kub{\'a}nek}, {Sota},
  {S{\'a}nchez-Ram{\'{\i}}rez}, {Andrei}, {Camargo}, {da Silva Neto}, {Gomes},
  {Martins}, {Gillon}, {Manfroid}, {Tozzi}, {Harlingten}, {Saravia}, {Behrend},
  {Mottola}, {Melendo}, {Peris}, {Fabregat}, {Madiedo}, {Cuesta}, {Eibe},
  {Ull{\'a}n}, {Organero}, {Pastor}, {de Los Reyes}, {Pedraz}, {Castro}, {de La
  Cueva}, {Muler}, {Steele}, {Cebri{\'a}n},
  {Monta{\~n}{\'e}s-Rodr{\'{\i}}guez}, {Oscoz}, {Weaver}, {Jacques}, {Corradi},
  {Santos}, {Reis}, {Milone}, {Emilio}, {Guti{\'e}rrez}, {V{\'a}zquez}, \&
  {Hern{\'a}ndez-Toledo}}]{Sicardy2011}
{Sicardy}, B., {Ortiz}, J.~L., {Assafin}, M., {et~al.} 2011, \nat, 478, 493

\bibitem[{{Souza} {et~al.}(2006){Souza}, {Babcock}, {Pasachoff}, {Gulbis},
  {Elliot}, {Person}, \& {Gangestad}}]{Souza2006}
{Souza}, S.~P., {Babcock}, B.~A., {Pasachoff}, J.~M., {et~al.} 2006, \pasp,
  118, 1550

\bibitem[{{Spectrum Instruments, Inc.}(2014)}]{SpectrumInstruments2014}
{Spectrum Instruments, Inc.} 2014, {Intelligent
  Reference/TM-4\texttrademark~GPS Time \& Frequency System - User Manual}

\bibitem[{{Spencer} {et~al.}(2006){Spencer}, {Stansberry}, {Grundy}, \&
  {Noll}}]{Spencer2006}
{Spencer}, J.~R., {Stansberry}, J.~A., {Grundy}, W.~M., \& {Noll}, K.~S. 2006,
  in {AAS/Division for Planetary Sciences Meeting Abstracts}, Vol.~38, 546

\bibitem[{{Stansberry} {et~al.}(2008){Stansberry}, {Grundy}, {Brown},
  {Cruikshank}, {Spencer}, {Trilling}, \& {Margot}}]{Stansberry2008}
{Stansberry}, J., {Grundy}, W., {Brown}, M., {et~al.} 2008, in {The Solar
  System Beyond Neptune}, ed. M.~A. {Barucci}, H.~{Boehnhardt}, D.~P.
  {Cruikshank}, A.~{Morbidelli}, \& R.~{Dotson} ({The University of Arizona
  Press}), 161--179

\bibitem[{{Stansberry} {et~al.}(2012){Stansberry}, {Grundy}, {Mueller},
  {Benecchi}, {Rieke}, {Noll}, {Buie}, {Levison}, {Porter}, \&
  {Roe}}]{Stansberry2012}
{Stansberry}, J., {Grundy}, W., {Mueller}, M., {et~al.} 2012, \icarus, 219, 676

\bibitem[{{Stern} {et~al.}(2015){Stern}, {Bagenal}, {Ennico}, {Gladstone},
  {Grundy}, {McKinnon}, {Moore}, {Olkin}, {Spencer}, {Weaver}, {Young}, {more},
  \& {even more}}]{Stern2015}
{Stern}, S.~A., {Bagenal}, F., {Ennico}, K., {et~al.} 2015, Science, 350

\bibitem[{{Stern} \& {Trafton}(2008)}]{Stern2008}
{Stern}, S.~A. \& {Trafton}, L.~M. 2008, in {The Solar System Beyond Neptune},
  ed. M.~A. {Barucci}, H.~{Boehnhardt}, D.~P. {Cruikshank}, A.~{Morbidelli}, \&
  R.~{Dotson} ({The University of Arizona Press}), 365--380

\bibitem[{{Szpak} {et~al.}(2015){Szpak}, {Chojnacki}, \& {van den
  Hengel}}]{Szpak2015}
{Szpak}, Z.~L., {Chojnacki}, W., \& {van den Hengel}, A. 2015, J. Math. Imaging
  Vision, 52, 173

\bibitem[{{Thirouin} {et~al.}(2014){Thirouin}, {Noll}, {Ortiz}, \&
  {Morales}}]{Thirouin2014}
{Thirouin}, A., {Noll}, K.~S., {Ortiz}, J.~L., \& {Morales}, N. 2014, \aap,
  569, A3

\bibitem[{{Timerson} {et~al.}(2013){Timerson}, {Brooks}, {Conard}, {Dunham},
  {Herald}, {Tolea}, \& {Marchis}}]{Timerson2013}
{Timerson}, B., {Brooks}, J., {Conard}, S., {et~al.} 2013, \planss, 87, 78

\bibitem[{{van Belle}(1999)}]{vanBelle1999}
{van Belle}, G.~T. 1999, \pasp, 111, pp. 1515

\bibitem[{{Vilenius} {et~al.}(2012){Vilenius}, {Kiss}, {Mommert}, {M\"uller},
  {Santos-Sanz}, {P\'al}, {Stansberry}, {Mueller}, {Peixinho}, {Fornasier},
  {Lellouch}, {Delsanti}, {Thirouin}, {Ortiz}, {Duffard}, {Perna}, {Szalai},
  {Protopapa}, {Henry}, {Hestroffer}, {Rengel}, {Dotto}, \&
  {Hartogh}}]{Vilenius2012}
{Vilenius}, E., {Kiss}, C., {Mommert}, M., {et~al.} 2012, \aap, 541, A94

\bibitem[{{Vilenius} {et~al.}(2014){Vilenius}, {Kiss}, {M\"uller}, {Mommert},
  {Santos-Sanz}, {P\'al}, {Stansberry}, {Mueller}, {Peixinho}, {Lellouch},
  {Fornasier}, {Delsanti}, {Thirouin}, {Ortiz}, {Duffard}, {Perna}, \&
  {Henry}}]{Vilenius2014}
{Vilenius}, E., {Kiss}, C., {M\"uller}, T., {et~al.} 2014, \aap, 564, A35

\bibitem[{{Wasserman} {et~al.}(1979){Wasserman}, {Millis}, {Franz}, {Bowell},
  {White}, {Giclas}, {Martin}, {Elliot}, {Dunham}, {Mink}, {Baron},
  {Honeycutt}, {Henden}, {Kephart}, {A'Hearn}, {Reitsema}, {Radick}, \&
  {Taylor}}]{Wasserman1979}
{Wasserman}, L.~H., {Millis}, R.~L., {Franz}, O.~G., {et~al.} 1979, \aj, 84,
  259

\bibitem[{{Wolf} {et~al.}(2014){Wolf}, {Wiedemann}, {Pf{\"u}ller},
  {Lachenmann}, {Hall}, \& {R{\"o}ser}}]{Wolf2014}
{Wolf}, J., {Wiedemann}, M., {Pf{\"u}ller}, E., {et~al.} 2014, \procspie, 9145,
  91450W

\bibitem[{{Wright} {et~al.}(2010){Wright}, {Eisenhardt}, {Mainzer}, {Ressler},
  {Cutri}, {Jarrett}, {Kirkpatrick}, {Padgett}, {McMillan}, {Skrutskie},
  {Stanford}, {Cohen}, {Walker}, {Mather}, {Leisawitz}, {Gautier}, {McLean},
  {Benford}, {Lonsdale}, {Blain}, {Mendez}, {Irace}, {Duval}, {Liu}, {Royer},
  {Heinrichsen}, {Howard}, {Shannon}, {Kendall}, {Walsh}, {Larsen}, {Cardon},
  {Schick}, {Schwalm}, {Abid}, {Fabinsky}, {Naes}, \& {Tsai}}]{Wright2010}
{Wright}, E.~L., {Eisenhardt}, P.~R.~M., {Mainzer}, A.~K., {et~al.} 2010, \aj,
  140, 1868

\end{thebibliography}

\appendix


\section{Density estimates} \label{app:densities}
Table~\ref{app:tab1} summarizes all TNOs for which a density estimate could be derived to this date. The size and mass of Pluto and its satellite Charon were measured during the New Horizons flyby. All remaining objects are binary systems, which allowed an estimate of the respective system mass based on the satellite's orbit. Only the size of Eris and Quaoar have been probed through observations of stellar occultations so far. All other size estimates were derived from thermophysical modeling of \textit{Spitzer} and/or \textit{Herschel} data. 

\begin{table*}[tbp]
  \setlength{\extrarowheight}{2.5pt}
  \centering
  \caption{Summary of TNOs with estimated bulk densities.}
  \label{app:tab1}
    \begin{tabular}{cccc}
    \hline\hline
    Target & Effective Diameter (km) & Bulk Density ($\mathrm{g~cm^{-3}}$) & Source \\     
    \hline
        (134340) Pluto  & $2374 \pm 8 $ & $1.860 \pm 0.013 $ & \citet{Stern2015} \\
        (136199) Eris & $2326 \pm 12$ & $2.52 \pm 0.05$ & \citet{Sicardy2011} \\
        (134340) I Charon & $1212 \pm 6$ & $1.702 \pm 0.021$ & \citet{Stern2015} \\
        (50000) Quaoar & $1110 \pm 5$ & $1.99 \pm 0.46$ & \citet{Braga-Ribas2013} \\
        (90482) Orcus\tablefootmark{a} & $958.4 \pm 22.9$ & $1.53^{+0.15}_{-0.13}$ & \citet{Fornasier2013} \\
        (120347) Salacia\tablefootmark{a} & $901 \pm 45$ & $1.29^{+0.29}_{-0.23}$ & \citet{Fornasier2013} \\
        (174567) Varda\tablefootmark{a} & $792^{+91}_{-84}$ & $1.27^{+0.41}_{-0.44}$ & \citet{Vilenius2014} \\
        (55637) 2002 UX\textsubscript{25}\tablefootmark{a} & $692 \pm 23$ & $0.82 \pm 0.11$ & \citet{Brown2013b} \\
        (47171) 1999 TC\textsubscript{36}\tablefootmark{a} & $393.1^{+25.2}_{-26.8}$ & $0.64^{+0.15}_{-0.11}$ & \citet{Mommert2012} \\ 
        (79360) Sila-Nunam\tablefootmark{a} & $343 \pm 42$ & $0.73 \pm 0.28$ & \citet{Vilenius2012} \\
        (148780) Altjira\tablefootmark{a} & $331^{+51}_{-187}$ & ${0.30}^{+0.50}_{-0.14}$ & \citet{Vilenius2014} \\
        2001 QC\textsubscript{298}\tablefootmark{a} & $303^{+27}_{-30}$ & $1.14^{+0.34}_{-0.30}$ & \citet{Vilenius2014} \\
        (26308) 1998 SM\textsubscript{165}\tablefootmark{a} & $279.8^{+29.7}_{-28.6}$ & $0.51^{+0.29}_{-0.14}$ & \citet{Stansberry2008}, \citet{Spencer2006} \\
        (65489) Ceto\tablefootmark{a}   & $281 \pm 11$  & $0.64^{+0.16}_{-0.13}$ & \citet{Santos-Sanz2012} \\
        (275809) 2001 QY\textsubscript{297}\tablefootmark{a} & $229^{+22}_{-108}$ & $0.92^{+1.30}_{-0.27}$ & \citet{Vilenius2014} \\
        2001 XR\textsubscript{254}\tablefootmark{a} & $221^{+41}_{-71}$ & $1.00^{+0.96}_{-0.56}$ & \citet{Vilenius2014} \\
        (88611) Teharonhiawako\tablefootmark{a} & $220^{+41}_{-44}$ & $0.60^{+0.36}_{-0.33}$ & \citet{Vilenius2014} \\
        (66652) Borasisi\tablefootmark{a} & $163^{+32}_{-66}$ & $2.1^{+2.60}_{-1.2}$ & \citet{Vilenius2014} \\
        (42355) Typhon\tablefootmark{a} & $157 \pm 34$ & $0.60^{+0.72}_{-0.29}$ & \citet{Stansberry2012} \\
    \hline
    \end{tabular}
\tablefoot{
\tablefoottext{a}{Binary system properties; primary not individually measured through occultation or spacecraft encounter.}
}
\end{table*}


\section{Error propagation} \label{app:errors}
Applying the laws of error propagation, the error of the albedo estimate can be calculated as follows:
\begin{equation}
\begin{split}
\Delta p = & \left[ \left( \frac{\mathrm{d} p}{\mathrm{d} H_\mathrm{Sun}} \Delta H_\mathrm{Sun}\right)^2 + \left( \frac{\mathrm{d} p}{\mathrm{d} H_\mathrm{TNO}} \Delta H_\mathrm{TNO}\right)^2 \right. \\ 
& + \left. \left( \frac{\mathrm{d} p}{\mathrm{d} a_\mathrm{El}} \Delta a_\mathrm{El}\right)^2 + \left( \frac{\mathrm{d} p}{\mathrm{d} b_\mathrm{El}} \Delta b_\mathrm{El}\right)^2 \right]^{1/2}
\end{split}
\label{app:eq1}
.\end{equation}
In our calculations we assumed $\Delta H_{Sun} = 0$. The remaining differentials are
\begin{equation}
\frac{\mathrm{d} p}{\mathrm{d} H_\mathrm{TNO}} = \frac{4 \cdot \left(1\,\mathrm{AU}\right)^2}{a_\mathrm{El} b_\mathrm{El}} 10^{0.4 H_\mathrm{Sun}} \ln{10} \cdot 10^{-0.4 H_\mathrm{TNO}}
\label{app:eq2}
\end{equation}

\begin{equation}
\frac{\mathrm{d} p}{\mathrm{d} a_\mathrm{El}} = - \frac{4 \cdot 10^{0.4 \left(H_\mathrm{Sun} - H_\mathrm{TNO}\right)}\cdot \left(1\,\mathrm{AU}\right)^2}{a_\mathrm{El}^2 b_\mathrm{El}^{}}
\label{app:eq3}
\end{equation}

\begin{equation}
\frac{\mathrm{d} p}{\mathrm{d} b_\mathrm{El}} = - \frac{4 \cdot 10^{0.4 \left(H_\mathrm{Sun} - H_\mathrm{TNO}\right)}\cdot \left(1\,\mathrm{AU}\right)^2}{a_\mathrm{El}^{} b_\mathrm{El}^2}
\label{app:eq4}
\end{equation}

\end{document}